\pdfoutput=1  
\def\acap{\\ \nonumber \\}

\def\rfr#1{Equation\,(\ref{#1})}
\def\rfrs#1#2{Equations\,(\ref{#1})--(\ref{#2})}
\def\Rfr#1{Equation\,(\ref{#1})}
\def\Rfrs#1#2{Equations\,(\ref{#1})--(\ref{#2})}
\def\derp#1#2{\rp{\partial{#1}}{\partial{#2}}}
\def\dert#1#2{\frac{{{\textrm{d}}}{#1}}{{{\textrm{d}}}{#2}}}
\def\virg#1{``#1"}
\def\eqi{\begin{equation}}
\def\eqf{\end{equation}}
\def\rp#1#2{\frac{#1}{#2}}
\def\lb#1{\label{#1}}

\def\ton#1{\left(#1\right)}
\def\qua#1{\left[#1\right]}
\def\grf#1{\left\{#1\right\}}
\def\ang#1{\left\langle #1\right\rangle}

\RequirePackage[2020-02-02]{latexrelease}
\documentclass{aastex631}
\usepackage{morefloats}
\usepackage[title]{appendix}
\usepackage{textcomp}
\usepackage{booktabs}
\usepackage{multirow}
\usepackage{rotating,tabularx}
\usepackage{float}
\usepackage{enumerate}
\usepackage{rotating}
\usepackage[polutonikogreek,english]{babel}
\usepackage{amsmath,starfont,textgreek,w-greek}
\usepackage[flushleft]{threeparttable}
\usepackage{amsthm}
\usepackage{bigints}
\usepackage{amscd}
\usepackage[mathlines]{lineno}
\usepackage{amssymb,dsfont}
\usepackage{graphicx,epsfig}
\usepackage{txfonts}
\usepackage{xr-hyper}
\usepackage{hyperref}
\usepackage[utf8]{inputenc}
\usepackage{newunicodechar,graphicx}

\usepackage[nointegrals]{wasysym}
\usepackage[caption=false]{subfig}

\allowdisplaybreaks

\makeatletter
\DeclareRobustCommand\ref{%
    \@ifstar\@refstar\T@ref
  }%
  \DeclareRobustCommand\pageref{%
    \@ifstar\@pagerefstar\T@pageref
  }%
 \makeatother

\begin{document}

\title{The Lense--Thirring effect at work in M87$^\ast$}

\shortauthors{L. Iorio}

\author[0000-0003-4949-2694]{Lorenzo Iorio}
\affiliation{Ministero dell' Istruzione e del Merito. Viale Unit\`{a} di Italia 68, I-70125, Bari (BA),
Italy}

\email{lorenzo.iorio@libero.it}

\begin{abstract}
Recently, the temporal evolution of the angles characterizing the spatial configuration of the jet in the supermassive black hole M87$^\ast$ was measured exhibiting a precessional pattern around the hole's spin axis. It would be due to the dragging induced by the fact that the hole's external spacetime is described by the Kerr metric.
Here, it is shown that the Lense--Thirring orbital precessions of a test particle moving about a rotating massive object, calculated perturbatively to the first post--Newtonian order, are able to fully reproduce all the measured features of the orientation of the jet axis of M87$^\ast$. In particular, by assuming that the latter is aligned with the  angular momentum of the accretion disk, modelled as an effective particle moving along a circular orbit, the condition that the absolute value of the predicted Lense--Thirring precessional frequency of the disk agrees with the measured value of $0.56\pm 0.02$ radians per year of the jet's one is satisfied for a range of physically meaningful values of the hole's spin parameter, close to unity, and of the effective disk radius, of the order of just over a dozen gravitational radii.
Relying upon such assumptions and results, it is possible to predict that the angle between the hole's spin axis and the jet's one stays constant over the years amounting to $1.16^\circ$, in agreement with its measured value of $1.25^\circ\pm 0.18^\circ$. Furthermore, also the temporal pattern and the amplitudes of the time series of the jet's angles are reproduced by the aforementioned Lense--Thirring precessional model.
\end{abstract}

\section{Introduction}
The focus of this paper is on the so--called Lense--Thirring (LT) effect\footnote{In fact, recent historical studies \cite{2007GReGr..39.1735P,2008mgm..conf.2456P,Pfister2014} pointed out that it would be more correct to rebrand it as Einstein--Thirring--Lense effect; however, the name which has now entered into common use will be used throughout the paper.} \cite{1918PhyZ...19..156L,1984GReGr..16..711M}, arising to the first post--Newtonian (1pN) order of the General Theory of Relativity (GTR), and its applicability to the context of the temporal evolution of the accretion disks surrounding supermassive black holes (SMBH) lurking at galactic cores. Emphasis will be given, in particular,  to the one at the centre of the\footnote{Also known as Virgo A or NGC 4486, where the acronym NGC stands for New General Catalogue, the eighty--seventh object of the Messier catalogue is a supergiant elliptical galaxy in the constellation Virgo about $16$ Megaparsec (Mpc) away from us that contains several trillion stars and hosts a  some billions of solar masses SMBH at its centre \cite{2019ApJ...875L...1E}. M87 was discovered by the French astronomer Charles Messier  in 1781. Its jet, the first to be discovered in its category, was reported in \cite{1918PLicO..13....9C}.}  M87 galaxy \cite{Berman15}, called M87$^\ast$ \cite{2013ARA&A..51..511K,2024MNRAS.527.2341S}, whose shadow was imaged with the Event Horizon Telescope (EHT) a few years ago \cite{2019ApJ...875L...1E}. Furthermore, a precessional motion of its jet \cite{2017Galax...5....2H} was recently measured by analyzing a record of radio observations collected with the Very Long Baseline Interferometry (VLBI) technique  over 22 years \cite{2023Natur.621..711C}. In \cite{2023Natur.621..711C}, on the basis of complicated general relativistic magnetohydrodynamic (GRMHD) simulations \cite{1997ApJ...476..221B,1999ApJ...522..727K,2003ApJ...599.1238D,2003ApJ...589..444G,2005PhRvD..72d4014S,2006EAS....21...43G,2013rehy.book.....R,2024arXiv240413824M}, it was suggested that just the LT effect may be a plausible candidate to explain it. A possible contribution of electric charge was recently studied in the framework of the Kerr--Newman metric \cite{1965JMP.....6..915N,1965JMP.....6..918N} as well  \cite{2024arXiv241107481M}.

This paper will confirm it clearly and transparently in a way hopefully understandable to a broader audience not specialized in GRMHD nor in black hole (BH) astrophysics.
The present work aims also to build a bridge that confidently unites, on the one hand,  the community that deals with celestial mechanics and astrodynamics and, on the other hand, the one that studies astrophysics of compact objects by translating the language used by one into that adopted by the other and vice versa. This should make it easier for the two communities to read each other's work, potentially stimulating new collaborations.

The following physical and orbital parameters are used throughout the paper.

$G$ is the Newtonian constant of universal gravitation, while $c$ is the speed of light in vacuum.

Furthermore, $M$ is the mass of a localized gravitational field source, whether it is a material body with a physical surface or not, such as a BH, acting as the primary in a  gravitationally bound restricted two--body system made of itself and a test particle orbiting it, $\upmu:=GM$, is its standard gravitational parameter, $R_\mathrm{g}:=\upmu/c^2$ is its gravitational radius, $\boldsymbol{J}$ is  its spin angular momentum, $J$ is its magnitude, and $\boldsymbol{\hat{k}}:=\boldsymbol{J}/J$ is its spin axis arbitrarily oriented with respect to some asymptotic\footnote{It means that, in principle, its inertial nature may be limited by some residual tidal effects due to external gravitational fields; here, they will be considered negligible.} inertial reference frame $\mathcal{K}$ centered in the field's source. For a rotating BH \cite{1970Natur.226...64B}, whose external spacetime is believed to be described by the Kerr metric \cite{1963PhRvL..11..237K,2015CQGra..32l4006T}, it is \cite{1986bhwd.book.....S} $J=\chi M^2 G/c$, where $\chi$ is a dimensionless number whose (absolute) value ranges from 0 (no rotation, static Schwarzshild BH) to\footnote{The aforementioned condition on the dimensionless spin parameter holds only for the Kerr metric. For material bodies, it is generally not satisfied. Suffice it to say that for, e.g., the Earth and Jupiter it is $\chi_\oplus\simeq 738,\chi_{\jupiter}\simeq 860$; for the Sun, it is $\chi_\odot\simeq 0.2$.} 1 (maximally rotating BH, or extreme Kerr BH). If $\left|\chi\right|>1$, a naked singularity \cite{1973CMaPh..34..135Y,1991PhRvL..66..994S} without a horizon would occur, implying the possibility of causality violations because of closed timelike curves. It may be worth of recalling that, although not yet proven, the cosmic censorship conjecture \cite{1999JApA...20..233P,2002GReGr..34.1141P} states that naked singularities may not be formed via the gravitational collapse of a material body. The parameter $\chi$, which can also be viewed as the second characteristic length\footnote{Sometimes, the symbol $a$ is used  for $J/\ton{Mc}$ itself, in which case it is dimensionally a length. } $J/\ton{Mc}$ occurring in the Kerr metric \cite{1963PhRvL..11..237K,2015CQGra..32l4006T} measured in units of the gravitational radius $R_\mathrm{g}$, is usually denoted with $a$ in BH studies.
The motion of an uncharged and nonspinning massive particle in the full Kerr metric was described in \cite{1972ApJ...178..347B,1972PhRvD...5..814W,2024PhRvD.109b4029A}. Ways to probe possible departures from the Kerr metric in the strong--field regime are discussed, e.g., in \cite{2024arXiv240602454G}.

As far as the orbit of the test particle is concerned, the semimajor axis is half the sum of its maximum and minimum distances  from the primary, so that it characterizes the orbit's size. In celestial mechanics, it is commonly denoted with $a$; thus, caution is in order since it may be confused with the BH's spin parameter.

The eccentricity, which determines the shape of the orbit,  is usually denoted with $e$, being $0\leq e<1$. A circle of radius $a\equiv r_0$  is obtained for $e=0$ (the maximum and the minimum distances from the primary are identical), while a highly eccentric orbit has $e$ close to unity. In BH studies, $e$ can reach very large values as for many S--type stars revolving around the SMBH in the Galactic Centre (GC) at Sagittarius A$^\ast$ (Sgr A$^\ast$) \cite{2009ApJ...692.1075G,2010RvMP...82.3121G,2017ApJ...837...30G}. On the other hand, if the test particle is representative of a fluid element of an accretion disk around a SMBH, $e$ is assumed to be zero.

The orientation of the orbital plane with respect to $\mathcal{K}$, or, equivalently, of the orbital angular momentum $\boldsymbol{h}$ of the test particle, is fixed by two angles which, in celestial mechanics, are customarily denoted with $I$ and $\mathit{\Omega}$. The former, called  inclination, is the tilt of $\boldsymbol{h}$ with respect to some direction of $\mathcal{K}$ chosen as reference $z$ axis, and ranges from $0^\circ$ to $180^\circ$. The orbits characterized by $0^\circ\leq I <  90^\circ$ are called  prograde, while those with $90^\circ < I \leq 180^\circ$ are defined as retrograde. Orbits with $I=90^\circ$ are said polar since they pass through the poles of the primary. In BH studies, $I$ is usually denoted with $\phi$ and called  viewing angle, while the reference $z$ direction is the line of sight (LOS), usually oriented towards the observer \cite{2023Natur.621..711C}. The other angle $\mathit{\Omega}$ is the longitude of the ascending node. It lies within the fundamental plane $\Pi$ of $\mathcal{K}$; the latter one is the reference plane  perpendicular to the direction with respect to which $I$ is reckoned; in general, it does not coincide with the primary's equatorial plane. $\mathit{\Omega}$, ranging from $0^\circ$ to $360^\circ$, is counted from the reference $x$ direction to the point on the line of nodes crossed by the particle from below, where the upward direction is that of $\boldsymbol{h}$; such a crossing point is the ascending node, denoted with $\ascnode$. In turn, the line of nodes is the intersection of the orbital plane with $\Pi$.
In the BH literature, $\Pi$ is the plane of the sky, while $\mathit{\Omega}$ coincides with the position angle (PA) of the sky--projected orbital angular momentum, being denoted with $\eta$ \cite{2023Natur.621..711C}, up to a constant offset of $90^\circ$ so that $\Omega = \eta + 90^\circ$.

Finally, the orientation of the orbit within the orbital plane itself, or, equivalently, of the line of apsides being the latter ones the pericentre and the apocentre, is determined by the argument of pericentre $\omega$. It is an angle which ranges from $0^\circ$ to $360^\circ$, and is reckoned within the orbital plane from the line of nodes towards the ascending node to the point of closest approach.

\begin{figure}
\centering
\begin{tabular}{c}
\includegraphics[width = 15 cm]{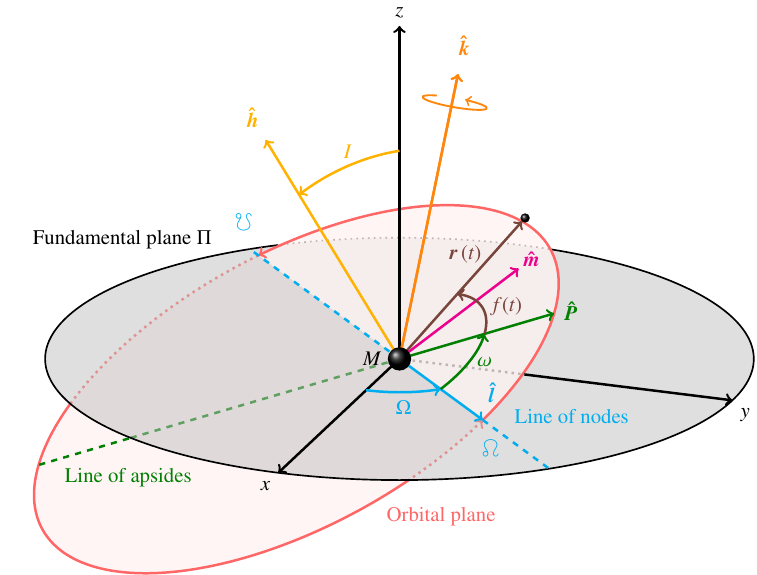}\\
\end{tabular}
\caption{Keplerian ellipse followed by a test particle orbiting an object of mass $M$ located at one of its foci at distance $r\ton{t}$. The fundamental plane $\Pi$ of the asymptotically inertial reference system $\mathcal{K}$ adopted is shaded in gray. The unit vector $\boldsymbol{\hat{k}}$ of the primary's spin angular momentum is arbitrarily oriented with respect to $\mathcal{K}$. The angles $I, \mathit{\Omega}$ and $\omega$, are the inclination, the longitude of the ascending node and the argument of pericentre, respectively.  The ascending and descending nodes are labelled with $\ascnode$ and $\descnode$, respectively.  The unit vectors $\boldsymbol{\hat{l}},\boldsymbol{\hat{m}},\boldsymbol{\hat{h}}$ and $\boldsymbol{\hat{P}}$, given by \rfrs{elle}{acca} and \rfr{Pi}, are clearly visible. The lines of nodes and apsides are shown dashed. Also the true anomaly $f\ton{t}$, reckoning the instantaneous position of the orbiter with respect to the pericentre, is displayed for completeness. }\label{figure0}
\end{figure}

\begin{figure}
\centering
\begin{tabular}{c}
\includegraphics[width = 15 cm]{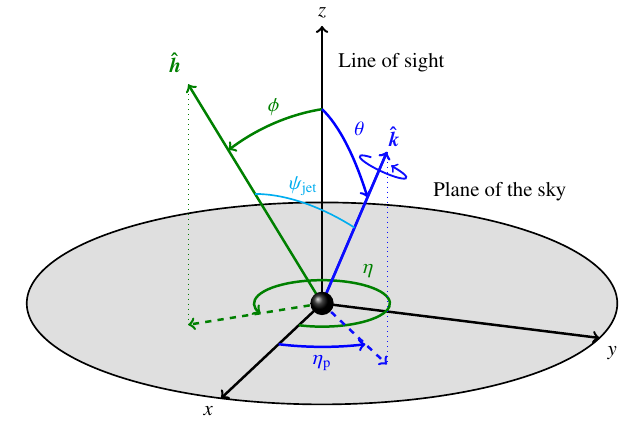}\\
\end{tabular}
\caption{SMBH's spin axis $\boldsymbol{\hat{k}}$ and unit vector of the disk's orbital angular momentum $\boldsymbol{\hat{h}}$. The notation of \cite{2023Natur.621..711C} is adopted for their angles whose numerical values were chosen here solely for illustrative purposes.}\label{figure0b}
\end{figure}

Figure \ref{figure0} displays a generic Keplerian ellipse, arbitrarily oriented in space and within its orbital plane, with a massive spinning primary located at one of its foci; its spin axis $\boldsymbol{\hat{k}}$ is not directed along any preferred direction.
Figure \ref{figure0b} is focussed on the SMBH's spin axis and the orbital angular momentum. For the sake of convenience, the notation of \cite{2023Natur.621..711C} is adopted for the angles determining their orientation in space.

If only the Newtonian mass monopole acceleration $A_\mathrm{N}=\upmu/r^2$ acts on the test particle, where $r$ is its instantaneous distance from its primary, all the aforementioned orbital parameters stay constant. Should some post--Keplerian (pK) accelerations $A_\mathrm{pK}$ other than $A_\mathrm{N}$ be present, altering the otherwise purely Keplerian motion, then (some of) the orbital parameters undergo steady temporal variations which manifest themselves cumulatively revolution after revolution.
Such a picture, which can be traced back to a perturbative scheme, is valid to the extent that the other pK accelerations are sufficiently smaller than $A_\mathrm{N}$.

The LT effect is a consequences of the so--called gravitomagnetic field arising from the off--diagonal components $g_{0i},\,i=1,2,3$ of the spacetime metric tensor $g_{\mu\nu},\,\mu,\nu=0,1,2,3$, which, in turn, are determined by the mass--energy currents of the source under consideration \cite{Thorne86,1986hmac.book..103T,1988nznf.conf..573T,2001rfg..conf..121M,2001rsgc.book.....R,Mash07}. For a localized rotating object, such as a star or a Kerr BH, the gravitomagnetic field is generated by its angular momentum $\boldsymbol{J}$.

Basically, the LT effect consists of variations of the orientation of both the orbital plane and of the orbit  within the orbital plane itself over time \cite{1918PhyZ...19..156L,1984GReGr..16..711M,2024gpno.book.....I}. Instead, the shape and the size of the path are left unaffected along with the time of passage at the pericentre.

Attempts to measure them in the Earth's field with some geodetic satellites \cite{2019JGeod..93.2181P} tracked with the Satellite Laser Ranging (SLT) technique \cite{SLR11} have been underway for almost 30 years \cite{1996NCimA.109..575C}. For different points of view on their reliability and actual accuracy, see, e.g., the reviews \cite{2011Ap&SS.331..351I,2013NuPhS.243..180C,2013CEJPh..11..531R}, and references therein. It should be said that the magnitude of the LT effect around the Earth amounts to a few tens of milliarcseconds per year (mas yr$^{-1}$), while it is expected to be as large as a few tens of sexagesimal degrees per year ($^\circ\,\mathrm{yr}^{-1}$)  around M87$^\ast$, i.e., more than a million times larger than around our planet. On the other hand, the several competing physical effects of non--gravitational origin which act as sources of systematic bias are generally much more accurately known in the terrestrial scenario than around SMBHs. Other tests of the LT effect were performed also in the fields of Jupiter \cite{2011AGUFM.P41B1620F,2024ApJ...971..145D} and the Sun \cite{2015IAUGA..2227771P,Pav2024,RussiLT}, but they proved to be inconclusive because of the resulting far too large errors and high correlations with other estimated parameters. The only known experiment that has successfully measured a gravitomagnetic effect and which, to date, has not been criticized in the peer--reviewed literature is that conducted by the Gravity Probe B (GP--B) mission \cite{Varenna74,2001LNP...562...52E}. Indeed, it detected, among other things, the gravitomagnetic Pugh--Schiff precessions \cite{Pugh59,Schiff60} of the spins of four gyroscopes carried onboard a drag--free spacecraft orbiting the Earth with a $19\%$ accuracy \cite{2011PhRvL.106v1101E,2015CQGra..32v4001E} compared to the $\simeq 1\%$ initially expected \cite{Varenna74,2001LNP...562...52E}.

The paper is organized as follows. Section \ref{sect:anal} reviews the analytical model of the LT effect, expressed in terms of the both the Keplerian orbital elements and in vectorial form, for a generic orientation of the primary's spin axis in space. In Section \ref{sec:M87}, the results of the previous Section are successfully applied to the case of the measured precession of the jet emanating from M87$^\ast$. Section \ref{fine} summarizes the findings and offers conclusions.

\section{Analytical modeling of the Lense--Thirring orbital precessions}\lb{sect:anal}
The LT rates of change of the inclination $I$, the longitude of the ascending node $\mathit{\Omega}$ and the argument of pericentre $\omega$, averaged\footnote{In the following, the angular brackets $\ang{\ldots}$ around the left-hand-sides of \rfrs{dotI}{doto}, usually denoting the orbital averages, are omitted in order to make the notation less cumbersome. } over one orbital revolution, can be analytically worked out with a variety of strategies; a straightforward and effective approach is a perturbative calculation made with the equations for the rates of change of the Keplerian orbital elements in Gaussian or Lagrange form \cite{Sof89,1991ercm.book.....B,2000ssd..book.....M,2003ASSL..293.....B,2005ormo.book.....R,2011rcms.book.....K,2014grav.book.....P,SoffelHan19,2024gpno.book.....I}. The resulting general expressions for the LT effect, valid for an arbitrary orientation of the primary's spin axis $\boldsymbol{\hat{k}}$, turn out to be \cite{2024gpno.book.....I}
\begin{align}
\dert I t \lb{dotI}& = \rp{2GJ\ton{\boldsymbol{\hat{k}}\boldsymbol\cdot\boldsymbol{\hat{l}}}}{c^2 a^3\ton{1 - e^2}^{3/2}},\acap
\dert{\mathit{\Omega}} t \lb{dotO}& = \rp{2GJ\csc I\ton{\boldsymbol{\hat{k}}\boldsymbol\cdot\boldsymbol{\hat{m}}}}{c^2 a^3\ton{1 - e^2}^{3/2}},\acap
\dert\omega t \lb{doto}& = -\rp{2GJ\qua{2\ton{\boldsymbol{\hat{k}}\boldsymbol\cdot\boldsymbol{\hat{h}}} + \cot I\ton{\boldsymbol{\hat{k}}\boldsymbol\cdot\boldsymbol{\hat{m}}}}}{c^2 a^3\ton{1 - e^2}^{3/2}},
\end{align}
where the unit vectors $\boldsymbol{\hat{l}}, \boldsymbol{\hat{m}}, \boldsymbol{\hat{h}}$ are defined as \cite{1991ercm.book.....B,Sof89,SoffelHan19}
\begin{align}
\boldsymbol{\hat{l}} \lb{elle} & = \grf{\cos\mathit{\Omega},\sin\mathit{\Omega},0},\acap
\boldsymbol{\hat{m}} \lb{emme} & = \grf{-\cos I \sin\mathit{\Omega},
\cos I \cos\Omega, \sin I},\acap
\boldsymbol{\hat{h}} \lb{acca} & = \grf{\sin I \sin\mathit{\Omega}, -\sin I \cos\mathit{\Omega}, \cos I}.
\end{align}
While $\boldsymbol{\hat{l}}$ is directed along the line of nodes towards the scending node $\ascnode$, $\boldsymbol{\hat{h}}$ is aligned with the orbital angular momentum in such a way that the relation
\eqi
\boldsymbol{\hat{l}}\boldsymbol\times\boldsymbol{\hat{m}} = \boldsymbol{\hat{h}}
\eqf
holds; cfr. with Figure \ref{figure0}.
\Rfrs{elle}{emme} allow to define the unit vector
\eqi
\boldsymbol{\hat{P}}:=\boldsymbol{\hat{l}}\cos\omega + \boldsymbol{\hat{m}}\sin\omega,\lb{Pi}
\eqf
which is directed within the orbital plane towards the pericentre: it determines the orientation of the Laplace--Runge--Lenz vector \cite{Gold80,Taff85}.
From \rfr{acca}, it explicitly turns out that, as anticipated,  $I$ and $\mathit{\Omega}$ define the orientation of the orbital angular momentum, or, equivalently, of the orbital plane, with respect to $\mathcal{K}$, while $\omega$ fixes the orientation of the orbit itself within the orbital plane.

\Rfrs{dotI}{doto} can be combined to yield  useful expressions for the rates of change of $\boldsymbol{\hat{h}}$ and $\boldsymbol{\hat{P}}$ as follows. Since they are functions of $I$, $\mathit{\Omega}$ and $\omega$,
%
their time derivatives can be written
\begin{align}
\dert{\boldsymbol{\hat{h}}}{t} \lb{dhdt}&= \derp{\boldsymbol{\hat{h}}}{I}\dert I t + \derp{\boldsymbol{\hat{h}}}{\mathit{\Omega}}\dert{\mathit{\Omega}}t,\acap
\dert{\boldsymbol{\hat{P}}}{t} \lb{dPdt}&= \derp{\boldsymbol{\hat{P}}}{I}\dert I t + \derp{\boldsymbol{\hat{P}}}{\mathit{\Omega}}\dert{\mathit{\Omega}}t + \derp{\boldsymbol{\hat{P}}}{\omega}\dert\omega t.
\end{align}
\Rfrs{dotI}{doto} allow to simultaneously put \rfrs{dhdt}{dPdt} in the compact form \cite{1974PhRvD..10.1340B}
\begin{align}
\dert{\boldsymbol{\hat{h}}}{t} \lb{vecprec}&= {\boldsymbol{\Omega}}^\mathrm{LT}\boldsymbol\times\boldsymbol{\hat{h}},\acap
\dert{\boldsymbol{\hat{P}}}{t} \lb{rlprec}&= {\boldsymbol{\Omega}}^\mathrm{LT}\boldsymbol\times\boldsymbol{\hat{P}},
\end{align}
where the precession velocity vector is defined as \cite{1974PhRvD..10.1340B}
\eqi
{\boldsymbol{\Omega}}^\mathrm{LT} := \rp{2G J}{c^2 a^3\ton{1 - e^2}^{3/2}}\qua{\boldsymbol{\hat{k}} - 3\ton{\boldsymbol{\hat{k}}\boldsymbol\cdot\boldsymbol{\hat{h}}}\boldsymbol{\hat{h}}}.\lb{Omeg}
\eqf
Indeed, a straightforward calculation shows that the right--hand--sides of \rfrs{vecprec}{rlprec}, calculated with \rfr{Omeg}, and of \rfrs{dhdt}{dPdt}, calculated with \rfrs{dotI}{doto}, are just identical.

The magnitude of the precession velocity vector ${\boldsymbol{\Omega}}^\mathrm{LT}$, given by \rfr{Omeg}, is
\eqi
\left|{\boldsymbol{\Omega}}^\mathrm{LT}\right| = \rp{2G J}{c^2 a^3\ton{1 - e^2}^{3/2}}\sqrt{1 + 3\ton{\boldsymbol{\hat{k}}\boldsymbol\cdot\boldsymbol{\hat{h}}}^2},
\eqf
while its orientation is given by
\eqi
{\boldsymbol{\hat{\Omega}}}^\mathrm{LT} = \rp{\boldsymbol{\hat{k}} - 3\ton{\boldsymbol{\hat{k}}\boldsymbol\cdot\boldsymbol{\hat{h}}}\boldsymbol{\hat{h}}}{\sqrt{1 + 3\ton{\boldsymbol{\hat{k}}\boldsymbol\cdot\boldsymbol{\hat{h}}}^2}}.
\eqf
\Rfrs{vecprec}{rlprec} tell that the orbital angular momentum and the Laplace--Runge--Lenz vector simultaneously precess about \rfr{Omeg}. However, if, as in the following, one is interested solely in $\boldsymbol{\hat{h}}$, a further precession velocity vector
\eqi
{\boldsymbol{\Omega}}^\mathrm{LT}_\mathrm{d} := \rp{2G J}{c^2 a^3\ton{1 - e^2}^{3/2}}\boldsymbol{\hat{k}}\lb{Omeg2}
\eqf
can be introduced. Indeed, it turns out that the right--hand--side of \rfr{vecprec} is the same if it is calculated with either \rfr{Omeg} or with \rfr{Omeg2}.

If $\upalpha$ is the angle between ${\boldsymbol{\Omega}}^\mathrm{LT}$ and $\boldsymbol{\hat{h}}$, one has
\begin{align}
\cos\upalpha \lb{cosalpha}& = {\boldsymbol{\hat{\Omega}}}^\mathrm{LT}\boldsymbol\cdot\boldsymbol{\hat{h}} = -\rp{2\ton{\boldsymbol{\hat{k}}\boldsymbol\cdot\boldsymbol{\hat{h}}}}{\sqrt{1 + 3\ton{\boldsymbol{\hat{k}}\boldsymbol\cdot\boldsymbol{\hat{h}}}^2}}, \acap
\sin\upalpha \lb{sinalpha} & = \left|{\boldsymbol{\hat{\Omega}}}^\mathrm{LT}\boldsymbol\times\boldsymbol{\hat{h}}\right| = \rp{\left|\boldsymbol{\hat{k}}\boldsymbol\times\boldsymbol{\hat{h}}\right|}{\sqrt{1 + 3\ton{\boldsymbol{\hat{k}}\boldsymbol\cdot\boldsymbol{\hat{h}}}^2}}.
\end{align}
According to \rfr{acca}, \rfrs{cosalpha}{sinalpha} are functions of $I$ and $\mathit{\Omega}$. Thus, it can be straightforwardly obtained
\eqi
\dert\upalpha t = \rp{1}{\cos\upalpha}\ton{\derp{\sin\upalpha}{I}\dert I t + \derp{\sin\upalpha}{\mathit{\Omega}}\dert{\mathit{\Omega}} t} = -\rp{1}{\sin\upalpha}\ton{\derp{\cos\upalpha}{I}\dert I t + \derp{\cos\upalpha}{\mathit{\Omega}}\dert{\mathit{\Omega}} t} = 0,\lb{dalphadt}
\eqf
where \rfrs{dotI}{dotO} were used.
\Rfr{dalphadt} proves explicitly that the motion of $\boldsymbol{\hat{h}}$ is just a precession about ${\boldsymbol{\hat{\Omega}}}^\mathrm{LT}$ since the angle $\upalpha$ between them remains constant. Note that in \rfr{dalphadt} it was assumed that $\boldsymbol{\hat{k}}$ is constant; such a condition will hold also in obtaining \rfr{dlambdadt} and \rfr{dbetadt}.

The same calculation can be repeated for the angle $\lambda$ between the unit vector of the precession velocity ${\boldsymbol{\hat{\Omega}}}^\mathrm{LT}$ and the primary's spin axis $\boldsymbol{\hat{k}}$. From
\eqi
\cos\lambda = {\boldsymbol{\hat{\Omega}}}^\mathrm{LT}\boldsymbol\cdot\boldsymbol{\hat{k}} = \rp{1 - 3\ton{\boldsymbol{\hat{k}}\boldsymbol\cdot\boldsymbol{\hat{h}}}^2}{\sqrt{1 + 3\ton{\boldsymbol{\hat{k}}\boldsymbol\cdot\boldsymbol{\hat{h}}}^2}},
\eqf
\eqi
\sin\lambda = \left|{\boldsymbol{\hat{\Omega}}}^\mathrm{LT}\boldsymbol\times\boldsymbol{\hat{k}}\right| =\rp{3\ton{\boldsymbol{\hat{k}}\boldsymbol\cdot\boldsymbol{\hat{h}}}\left|\boldsymbol{\hat{k}}\boldsymbol\times\boldsymbol{\hat{h}}\right|}{\sqrt{1 + 3\ton{\boldsymbol{\hat{k}}\boldsymbol\cdot\boldsymbol{\hat{h}}}^2}},
\eqf
it turns out that
\eqi
\dert\lambda t = \rp{1}{\cos\lambda}\ton{\derp{\sin\lambda}{I}\dert I t + \derp{\sin\lambda}{\mathit{\Omega}}\dert{\mathit{\Omega}} t} = -\rp{1}{\sin\lambda}\ton{\derp{\cos\lambda}{I}\dert I t + \derp{\cos\lambda}{\mathit{\Omega}}\dert{\mathit{\Omega}} t} = 0,\lb{dlambdadt}
\eqf
where \rfrs{dotI}{dotO} were used as again. Also $\lambda$ stays constant.

Let $\beta$ be the angle between the primary's spin axis $\boldsymbol{\hat{k}}$ and the unit vector of the orbital angular momentum $\boldsymbol{\hat{h}}$.
Then, from
\begin{align}
\cos\beta & = \boldsymbol{\hat{k}}\boldsymbol\cdot\boldsymbol{\hat{h}},\acap
\sin\beta & = \left|\boldsymbol{\hat{k}}\boldsymbol\times\boldsymbol{\hat{h}}\right|,
\end{align}
and \rfrs{dotI}{dotO},
it turns out that
\eqi
\dert\beta t = \rp{1}{\cos\beta}\ton{\derp{\sin\beta}{I}\dert I t + \derp{\sin\beta}{\mathit{\Omega}}\dert{\mathit{\Omega}} t} = -\rp{1}{\sin\beta}\ton{\derp{\cos\beta}{I}\dert I t + \derp{\cos\beta}{\mathit{\Omega}}\dert{\mathit{\Omega}} t} = 0.\lb{dbetadt}
\eqf
Also $\beta$ is constant.

Numerical integrations, to be discussed in the next Section, confirmed \rfr{dalphadt}, \rfr{dlambdadt} and \rfr{dbetadt}.
\section{Application to the accretion disk of M87$^\ast$}\lb{sec:M87}
Here, the analytical results of the previous Section are successfully applied to the case of the accretion disk around M87$^\ast$ in explaining its recently measured dynamical features \cite{2023Natur.621..711C}. In particular, the value of the angle between the SMBH's spin axis and the jet axis, assumed coincident with the disk's orbital angular momentum, predicted with the present LT model is in agreement with its measured counterpart \cite{2023Natur.621..711C}. Furthermore, the LT temporal evolution of the angles characterizing the orientation in space of the jet axis, calculated with \rfrs{dotI}{dotO}, reproduces that measured in \cite{2023Natur.621..711C}.

First of all, \rfrs{dotI}{dotO}, or, equivalently, \rfr{vecprec} and \rfr{Omeg2}, are able to immediately explain why an accretion disk does not precess if it is aligned with the equatorial plane of its host SMBH. Indeed, in this case
\begin{align}
\boldsymbol{\hat{k}}\boldsymbol\cdot\boldsymbol{\hat{l}} &= 0, \acap
\boldsymbol{\hat{k}}\boldsymbol\cdot\boldsymbol{\hat{m}} &= 0, \acap
\boldsymbol{\hat{k}}\boldsymbol\cdot\boldsymbol{\hat{h}} &=\pm 1,\acap
\boldsymbol{\hat{k}}\boldsymbol\times\boldsymbol{\hat{h}} &=0
\end{align}
hold. Thus, the rates of change of the inclination and the longitude of the ascending node of a fluid element of the disk, which, in this case takes the place of the test particle, are zero, as per \rfrs{dotI}{dotO}. Furthermore, since ${{\boldsymbol{\Omega}}^\mathrm{LT}}_\mathrm{d}$ is aligned with $\boldsymbol{\hat{h}}$, the latter stays constant according to \rfr{vecprec} and \rfr{Omeg2}. If, instead, the disk is misaligned with respect to the SMBH's equator, i.e., if it is
\begin{align}
\left|\boldsymbol{\hat{k}}\boldsymbol\cdot\boldsymbol{\hat{h}}\right|\lb{noalign}&\neq 1,\acap
\boldsymbol{\hat{k}}\boldsymbol\times\boldsymbol{\hat{h}}\lb{noalign2}&\neq 0,
\end{align}
then, according to \rfrs{dotI}{dotO}, the disk generally undergoes a precession since the SMBH's spin axis is not perpendicular to $\boldsymbol{\hat{l}}$ and $\boldsymbol{\hat{m}}$. Moreover, since the disk's orbital angular momentum and the SMBH's axis are not aligned, as per \rfrs{noalign}{noalign2}, \rfr{vecprec} and \rfr{Omeg2} tell that the rate of change of $\boldsymbol{\hat{h}}$ does not vanish.

The main assumptions on which the present analysis is based are the following ones.
\begin{itemize}
\item The spin axis $\boldsymbol{\hat{k}}$ of M87$^\ast$ is assumed to be constant, as done in \cite{2023Natur.621..711C}.
    In realistic astrophysical environments, such a feature is generally thought to be effectively constant within the timescales relevant to accretion disk and jet precession. However, it should be noted that, from a theoretical point of view, the BH's spin axis may undergo temporal variations under specific conditions, though such changes would be very gradual and rare.
    Here, the spin axis of M87$^\ast$,  which is the blue vector in Figure 2 $\boldsymbol{\ton{\mathrm{c}}}$ of \cite{2023Natur.621..711C} where it is called precession axis, is parameterized as
    \eqi
    \boldsymbol{\hat{k}}=\grf{\sin\theta \sin\eta_\mathrm{p}, -\sin \theta \cos\eta_\mathrm{p}, \cos \theta},\lb{kappa}
    \eqf
    where \cite{2023Natur.621..711C}
    \begin{align}
    \theta \lb{theta}&= 17.21^\circ, \acap
    \eta_\mathrm{p} \lb{etap}& = 288.47^\circ
    \end{align}
\Rfr{kappa}, calculated with \rfrs{theta}{etap}, yields
\eqi
\boldsymbol{\hat{k}} = \grf{-0.28, -0.09, 0.95},
\eqf
which is in visual agreement with Figure 2 $\boldsymbol{\ton{\mathrm{c}}}$ of \cite{2023Natur.621..711C}; indeed, in it, the blue vector has a large $z$ component, while both the $x$ and $y$ components are negative. Of these, the $x$ component appears to be much longer than the $y$ component.
\item The jet axis, which is the green vector in Figure 2 $\boldsymbol{\ton{\mathrm{c}}}$ of  \cite{2023Natur.621..711C}, is assumed essentially coincident with the unit vector $\boldsymbol{\hat{h}}$ of the disk's orbital angular momentum. Disk--jet tight coupling is supported for  tilted thick accretion discs around rapidly spinning BHs by extensive 3D GRMHD simulations \cite{2013Sci...339...49M,2018MNRAS.474L..81L}; see also \cite{2024arXiv241010965C}. An efficient disk--jet coupling was also assumed for OJ287 \cite{2023ApJ...951..106B} and AT2020ocn \cite{2024Natur.630..325P}. On the other hand, the alignment between the jet  and the accretion disk's orbital angular momentum  should not be deemed as absolutely perfect. Turbulence, magnetic instabilities, or plasma interactions within the jet can cause small deviations in its orientation relative to the disk's angular momentum. Furthermore, it depends on the actual process of jet launching. The Blandford--Payne (BP) process \cite{1982MNRAS.199..883B}
is connected to the disc structure, while the Blandford--Znajek (BZ) one \cite{1977MNRAS.179..433B} is related to the SMBH's ergosphere.
In principle, this could imply that in many cases the BP process is more applicable to large--scale jets,
but caution should be taken from source to source. \textcolor{black}{In principle, if the disk is fully saturated with magnetic flux, i.e. if it is in a magnetically arrested state, it may be difficult to justifiy a tight jet-disk alignment yielding a rigid precession \cite{2024ApJ...973..141G,2023Galax..11....4A}. Indeed, in this case, GRMHD simulations show that the torques in the system tend to align the inner disk with the jet (see, e.g., \cite{2013Sci...339...49M}). A certain degree of alignment may occur also in not magnetically saturated systems, as in the so-called standard and normal evolution (SANE) models in the GRMHD literature \cite{2024ApJ...974..209G}. Be that as it may, in general, it is easier to align more mildly tilted disks. Remarkably,} the results of this work strongly support the hypothesis of a tight coupling between $\boldsymbol{\hat{k}}$ and $\boldsymbol{\hat{h}}$ for M87$^\ast$, as it will be shown in the following.
\item In order to make a quantitative comparison with the experimental results of  \cite{2023Natur.621..711C}, a sort of \virg{effective} disk radius $r_0$, which is somehow representative of the global disk precession assumed rigid, is used in \rfrs{dotI}{dotO}. A similar approach was followed by, e.g., \cite{1998ApJ...492L..59S,1999ApJ...524L..63S,2009MNRAS.397L.101I} to explain the phenomenon of Quasi--Periodic Oscillations (QPOs) in the X--ray emission from accreting stellar mass BHs and neutron stars \cite{2016AN....337..398M}.
\textcolor{black}{Constraining the size of $r_0$ is of crucial importance to shed light on the intricate mechanisms underlying the disk's accretion and formation. A small effective radius of the disk may suggest that the torus accretion timescale is relatively short, such that its structure, and, thus the emitted radiation, qualitatively changes over time intervals decades long. On the other hand, such an issue is escapable if the disk is not too compact and the accretion process is not too efficient. Further, the effective radius of such a system should be set by the circularization radius of the infalling gas forming the disk, which is in general much larger than the event horizon. Recently, \cite{2024ApJ...973..141G} simulated the accretion flow from the M87 galaxy to the SMBH, wherein the radial size of the system is orders of magnitude larger than the event horizon.}
\end{itemize}

\begin{figure}
\centering
\begin{tabular}{c}
\includegraphics[width = 10 cm]{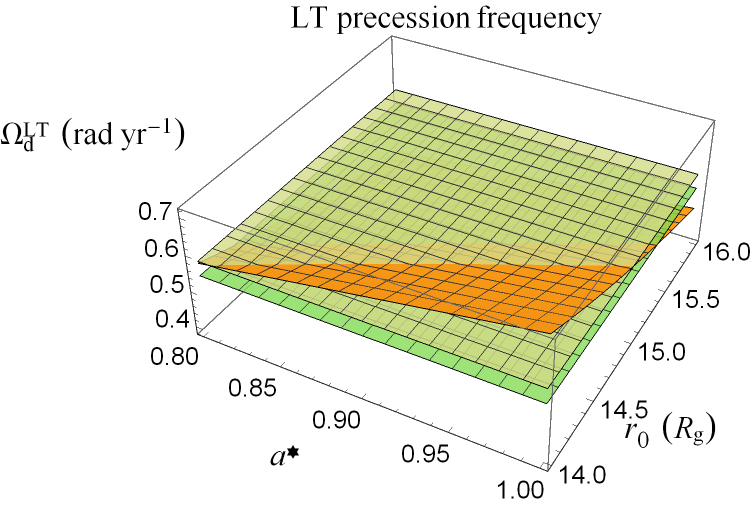}\\
\includegraphics[width = 10 cm]{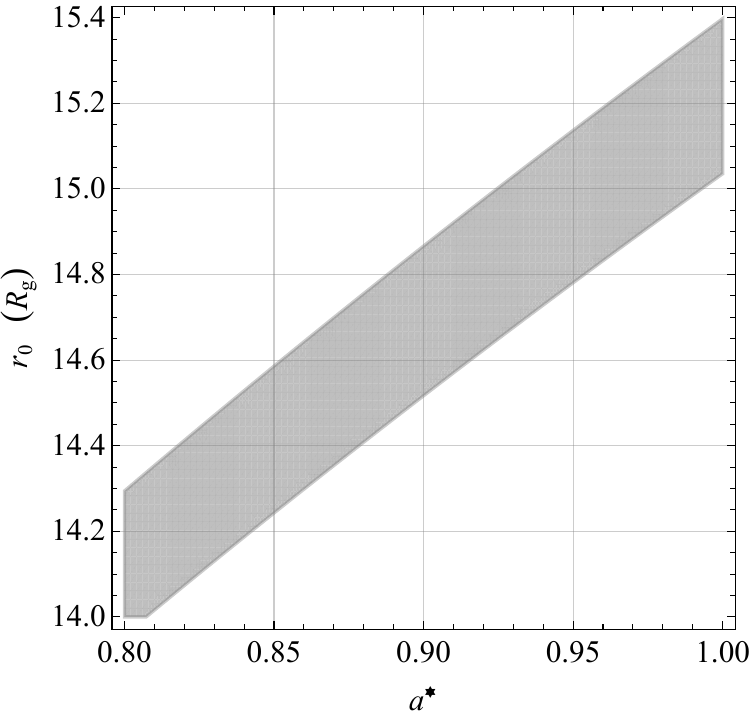}\\
\end{tabular}
\caption{Upper panel: plot of $\Omega^\mathrm{LT}_\mathrm{d}\ton{a^\ast,r_0}$ as given by \rfr{Omeg2}, in rad yr$^{-1}$, as a function of $a^\ast >0$ and $r_0$.
The upper and lower experimental values of $\left|\Omega_\mathrm{d}^\mathrm{exp}\right|\equiv\left|\omega^\mathrm{exp}_\mathrm{p}\right|$ as per \cite{2023Natur.621..711C} are depicted as  horizontal constant surfaces as well.   Lower panel: allowed region, in gray, in the $\grf{a^\ast,r_0}$ plane corresponding to the condition that the graph of $\Omega_\mathrm{d}^\mathrm{LT}\ton{a^\ast,r_0}$ of the upper panel remains confined between the upper and lower experimentally allowed values of $\left|\omega_\mathrm{p}^\mathrm{exp}\right|$, i.e. for $0.54\,\mathrm{rad\,yr}^{-1}\leq\Omega^\mathrm{LT}_\mathrm{d}\ton{a^\ast,r_0}\leq 0.58\,\mathrm{rad\,yr}^{-1}$. The value $e=0.9375$ adopted in \cite{2023Natur.621..711C}, based on \cite{2004ApJ...602..312G}, falls within the depicted permitted region, corresponding to $r_0\simeq 14.9\pm 0.2\,R_\mathrm{g}$. If $\left|\Omega\ton{a^\ast,r_0}\right|$ for $-1\leq a^\ast\leq 1$ were considered, also a second, identical branch curving to the left and symmetric to the $a^\ast=0$ axis would occur; cfr. with Figure 1 of \cite{2024arXiv241010965C}
}\label{figure:allow}
\end{figure}

\begin{figure}
\centering
\begin{tabular}{c}
\includegraphics[width = 10 cm]{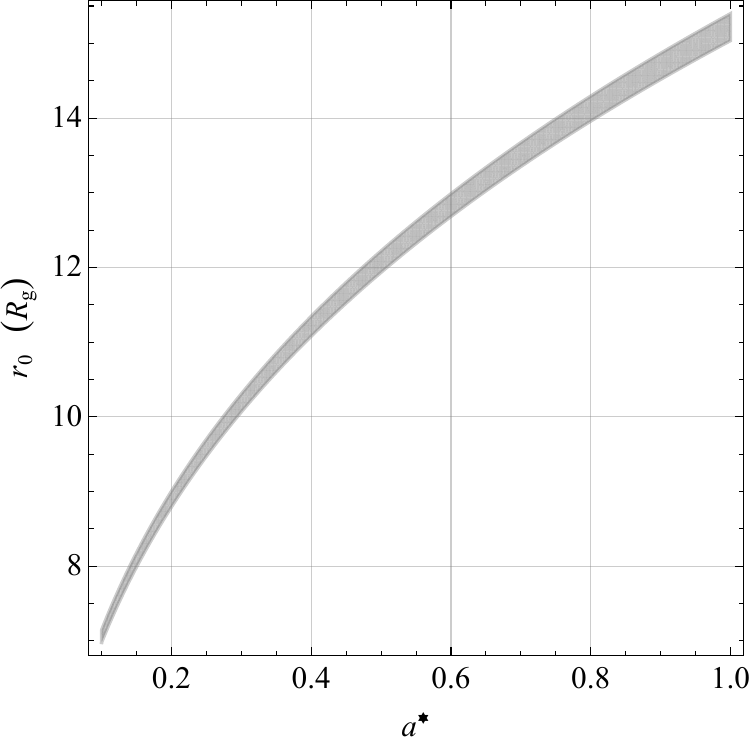}\\
\end{tabular}
\caption{Entire allowed region, in gray, in the $\grf{a^\ast,r_0}$ plane, with $a^\ast>0$, corresponding to the condition that the graph of $\Omega_\mathrm{d}^\mathrm{LT}\ton{a^\ast,r_0}$ remains confined between the upper and lower experimentally allowed values of $\left|\omega_\mathrm{p}^\mathrm{exp}\right|$, i.e., $0.54\,\mathrm{rad\,yr}^{-1}\leq\Omega^\mathrm{LT}_\mathrm{d}\ton{a^\ast,r_0}\leq 0.58\,\mathrm{rad\,yr}^{-1}$. If $\left|\Omega\ton{a^\ast,r_0}\right|$ for $-1\leq a^\ast\leq 1$ were considered, also a second, identical branch curving to the left and symmetric to the $a^\ast=0$ axis would occur; cfr. with Figure 1 of \cite{2024arXiv241010965C}.
}\label{figure:allow2}
\end{figure}

The analytical expression of the LT velocity precession $\Omega^\mathrm{LT}_\mathrm{d}$ of $\boldsymbol{\hat{h}}$ about the SMBH's spin axis $\boldsymbol{\hat{k}}$, given by \rfr{Omeg2}, can be viewed as a function of two independent variables: the spin parameter $a^\ast$ of M87$^\ast$ and the effective LT radius $r_0$ of a circular matter ring representative of its accretion disk. Thus, \rfr{Omeg2}, combined with  the experimental range of the measured values \cite[Tab.\,1]{2023Natur.621..711C} for the jet precession velocity, called $\omega_\mathrm{p}$ in \cite{2023Natur.621..711C},
\eqi
\left|\omega^\mathrm{exp}_\mathrm{p}\right| = 0.56\pm 0.02\,\mathrm{rad\,yr}^{-1},
\eqf
allows to obtain an exclusion plot in the  $\grf{a^\ast,r_0}$ plane. Indeed, by imposing the condition
\eqi
0.54\,\mathrm{rad\,yr}^{-1}\leq \Omega_\mathrm{d}^\mathrm{LT}\ton{a^\ast,r_0}\leq 0.58\,\mathrm{rad\,yr}^{-1},\lb{condiz}
\eqf
it is possible to exclude those values of $a^\ast >0$ and $r_0$ for which \rfr{condiz} is not fulfilled.  By taking into account the fact that several techniques have shown so far that the spin parameter of M87$^\ast$ is likely close to unity or, in any case, well greater than $\simeq 0.1$ \cite{1998AJ....116.2237K,2008ApJ...676L.109W,2009ApJ...699..513L,2012Sci...338..355D,2017MNRAS.470..612F,2018MNRAS.479L..65S,2019ApJ...880L..26N,2019MNRAS.489.1197N,2020AnP...53200480G,2020MNRAS.492L..22T,2023Astro...2..141D}, one finally gets the permitted region depicted in\footnote{Identical, symmetric branches would occur in Figures \ref{figure:allow} to \ref{figure:allow2} if $\left|\Omega_\mathrm{d}^\mathrm{LT}\right|$ and also negative values of $a^\ast$ were considered; see Figure 1 of \cite{2024arXiv241010965C}.} Figure \ref{figure:allow}.
It shows that there is a nearly linear region of allowed values  for $a^\ast$ and $r_0$ in the domain $0.8\leq a^\ast\leq 1\times 14\,R_\mathrm{g}\leq r_0\leq 15.4\,R_\mathrm{g}$. For any given value of $a^\ast$  ranging from $0.80$ to 1, there is a narrow set of permitted orbital radii about  $0.2\,R_\mathrm{g}$ wide. It should be noted that $a^\ast = 0.9375$, adopted in \cite{2023Natur.621..711C} on the basis of \cite{2004ApJ...602..312G}, falls well within the allowed region of Figure \ref{figure:allow}, corresponding to $14.7\,R_\mathrm{g}\lesssim r_0\lesssim 15.1\,R_\mathrm{g}$. Remarkably, such values for the effective disk radii are in agreement with the figure $\simeq 15\,R_\mathrm{g}$ given in \cite{2023Natur.621..711C}.
Even for such small values of the effective radius,  \rfrs{dotI}{doto}, obtained perturbatively to the 1pN order, are quite adequate for the case at hand. Indeed, the LT acceleration, calculated for $r_0=15\,R_\mathrm{g}$, amounts to no more than $2\%$ of the dominant Newtonian  one.
Figure \ref{figure:allow2}, which agrees with Figure 8 of \cite{2024PhRvD.110f4006W} and Figure 1 of \cite{2024arXiv241010965C}, shows that, in fact, other allowed regions exist in the $\grf{a^\ast,r_0}$ plane. Actually, they correspond to values of the spin parameter which are generally incompatible with the majority of the  constraints  published in the literature.
As far as the physically meaningful minimum value of $r_0$ is concerned, it is reasonable to assume the radius $r_\mathrm{ISCO}$ of the innermost stable circular orbit (ISCO)  for it. For a nonspinning BH, it is equal to $6 R_\mathrm{g}$, while for a circular tilted orbit  around a Kerr BH it depends on $a^\ast$ and $\theta$ \cite{2024PhRvD.109b4029A}. For $\theta\simeq 0^\circ$ and $\left|a^\ast\right|\lesssim 1$, the radius of the tilted innermost stable circular orbit (TISCO) is $r_\mathrm{TISCO}\gtrsim 1 R_\mathrm{g}\,\ton{a^\ast>0}$ or $r_\mathrm{TISCO}\lesssim 9 R_\mathrm{g}\,\ton{a^\ast < 0}$, as per Figure 1 of \cite{2024PhRvD.109b4029A}.

From the point of view of the corroboration of the hypothesis that the LT effect is  responsible for the phenomenology measured in \cite{2023Natur.621..711C}, it is certainly important that \rfr{condiz} is actually satisfied by physically plausible values of the spin parameter of M87$^\ast$ and of $r_0$. Nonetheless, it must be also demonstrated that, for such allowed values of the parameter space considered, the LT effect is able to accommodate other measured dynamical features of the jet--M87$^\ast$ system.

\begin{figure}
\centering
\begin{tabular}{c}
\includegraphics[width = 15 cm]{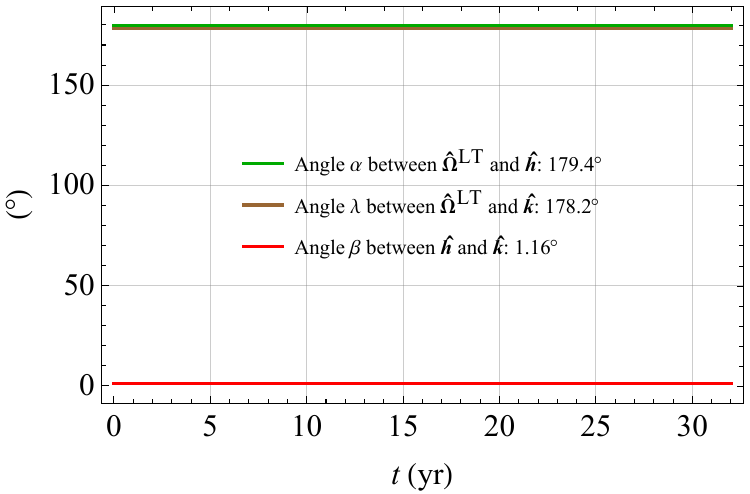}\\
\includegraphics[width = 15 cm]{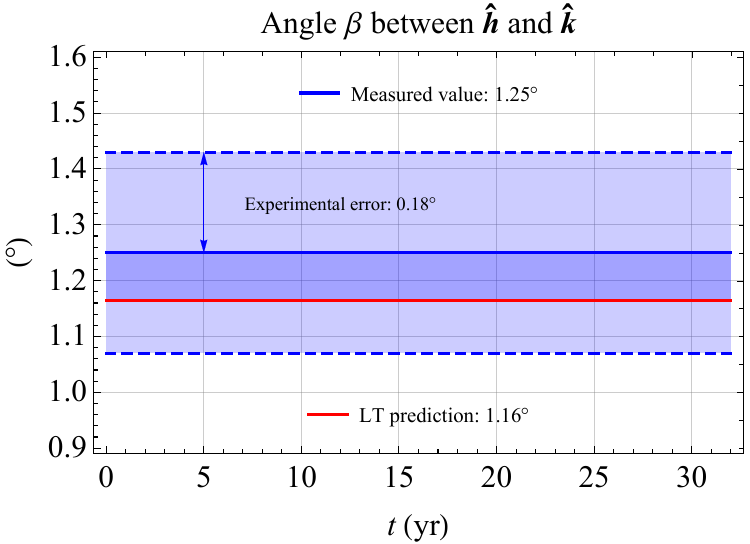}\\
\end{tabular}
\caption{Upper panel: numerically produced LT time series, in $^\circ$, for the angles $\alpha,\lambda$ and $\beta$ between ${\boldsymbol{\hat{\Omega}}}^\mathrm{LT}, \boldsymbol{\hat{h}}$ and $\boldsymbol{\hat{k}}$. Lower panel: numerically produced LT time series, in $^\circ$, for the angle  between $\boldsymbol{\hat{h}}$ and $\boldsymbol{\hat{k}}$ and the experimental range for it according to Table 1 of \cite{2023Natur.621..711C}. The values $a^\ast=0.9375, r_0=14.9\,R_g = 14.9\,\upmu_\bullet/c^2, M_\bullet = 6.5\times 10^9 M_\odot$ are used for the simultaneous numerical integration of \rfrs{dotI}{dotO} along with $I_0 = 17.85^\circ,\mathit{\Omega}_0 = 291.7^\circ$, retrieved from  Figure 2 $\boldsymbol{\ton{\mathrm{b}}}$ and Extended Data Figure 4 of  \cite{2023Natur.621..711C}, for the initial conditions of $I$ and $\mathit{\Omega}$.
}\label{figure:spin}
\end{figure}

To this aim, the equations for the averaged rates of change of the inclination $I$ and the longitude of the ascending node $\mathit{\Omega}$, which correspond to the angles $\phi$ and $\eta$, respectively, of \cite{2023Natur.621..711C}, are numerically integrated over, say, 50 yr according to \rfrs{dotI}{dotO} by using
\begin{align}
a^\ast & = 0.9375, \acap
r_0 &= 14.9\,R_\mathrm{g}
\end{align}
for the SMBH's spin parameter and the disk's effective radius,
and
\begin{align}
I_0 & = 17.85^\circ,\acap
\mathit{\Omega}_0 & = 291.7^\circ
\end{align}
for the initial values of $I$ and $\mathit{\Omega}$ retrieved from Figure 2 $\boldsymbol{\ton{\mathrm{b}}}$ and Extended Data Figure 4 of  \cite{2023Natur.621..711C}. As far the mass of M87$^\ast$ is concerned, the value \cite{2019ApJ...875L...1E}
\eqi
M=6.5\times 10^9 M_\odot\eqf
is adopted, where $M_\odot$ is mass of the Sun.
The upper panel of Figure \ref{figure:spin} displays the resulting time series of the angles $\alpha,\lambda$ and $\beta$, thus confirming their constancy, as per \rfr{dalphadt}, \rfr{dlambdadt} and \rfr{dbetadt}. The bottom panel of Figure \ref{figure:spin} shows the time series of the angle $\beta$, which corresponds to $\psi_\mathrm{jet}$ of \cite{2023Natur.621..711C}, superimposed to the experimental range for its measurement reported in \cite{2023Natur.621..711C}.
It turns out that its theoretical value
\eqi
\beta^\mathrm{LT}\equiv\psi^\mathrm{LT}_\mathrm{jet}=1.16^\circ,
\eqf
predicted according to the present model based on the LT effect, is in agreement with its measured counterpart
\eqi
\psi^\mathrm{exp}_\mathrm{jet} = 1.25^\circ\pm 0.18^\circ,
\eqf
reported in Table 1 of \cite{2023Natur.621..711C}. Incidentally, the predicted LT values of the angles $\alpha$ and $\lambda$ amount to $179.4^\circ$ and $178.2^\circ$, respectively.

\begin{figure}
\centering
\begin{tabular}{c}
\includegraphics[width = 15 cm]{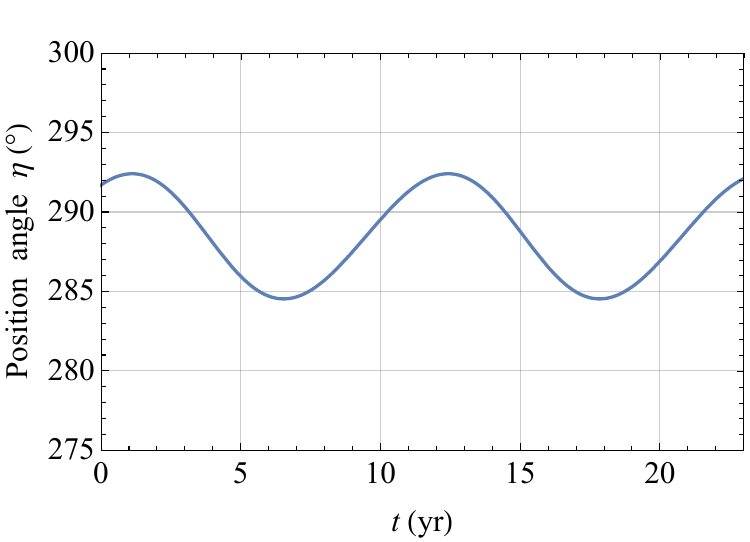}\\
\includegraphics[width = 15 cm]{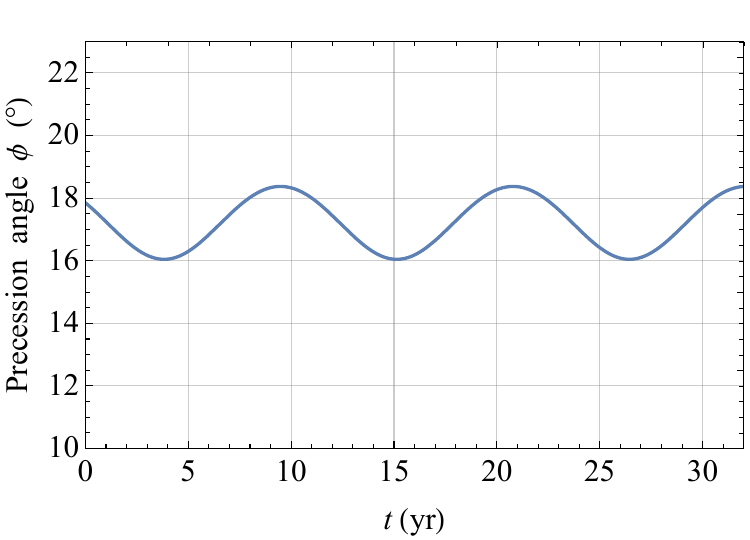}\\
\end{tabular}
\caption{Upper panel: numerically produced LT time series, in $^\circ$, for the angle $\eta$. Lower panel: numerically produced LT time series, in $^\circ$, for the angle $\phi$. The values $a^\ast=0.9375, r_0=14.9\,R_g = 14.9\,\upmu_\bullet/c^2, M_\bullet = 6.5\times 10^9 M_\odot$ are used for the simultaneous numerical integration of \rfrs{dotI}{dotO} along with $\phi_0 = 17.85^\circ,\eta_0 = 291.7^\circ$, retrieved from  Figure 2 $\boldsymbol{\ton{\mathrm{b}}}$ and Extended Data Figure 4 of  \cite{2023Natur.621..711C}, for the initial conditions of $\phi$ and $\eta$.
The time spans and the ranges of values on the vertical axes of both panels have the same length of those in Figure 2  and Extended Data Figure 4 of  \cite{2023Natur.621..711C} for a better comparison with the latter ones.}\label{figure:etaphi}
\end{figure}

Furthermore, the numerically produced time series for $I\ton{t}$ and $\mathit{\Omega}\ton{t}$, or, equivalently, for $\phi\ton{t}$ and $\eta\ton{t}$ in the notation of \cite{2023Natur.621..711C}, obtained with the previous integration of \rfrs{dotI}{dotO} and displayed in Figure \ref{figure:etaphi}, agree with those experimentally determined in
\cite{2023Natur.621..711C} and shown in their Figure 2 $\boldsymbol{\ton{\mathrm{b}}}$ $\ton{\eta}$ and Extended Data Figure 4 $\ton{\phi}$.
%
%
%
%
%
%
\section{Summary and conclusions}\lb{fine}
It was explicitly shown that the LT effect for the motion of a test particle following a circular orbit of radius $r_0$, written for an arbitrary orientation of the primary's spin axis $\boldsymbol{\hat{k}}$, is able to reproduce all the recently measured dynamical features of the jet precession of M87$^\ast$, assumed aligned with the accretion disk's orbital angular momentum $\boldsymbol{\hat{h}}$, for a value of the SMBH's dimensionless spin parameter as large as $a^\ast=0.9375$ and an effective radius $r_0$ amounting to just over a dozen gravitational radii.

More specifically, such values lie within the allowed region in the $\grf{a^\ast,r_0}$ plane obtained by imposing that the LT prediction $\left|\Omega_\mathrm{d}^\mathrm{LT}\right|$ for the precession velocity of the jet, viewed as a function of $a^\ast$ and $r_0$, agrees with its measured value $\left|\omega^\mathrm{exp}_\mathrm{p}\right|=0.56\pm 0.02\,\mathrm{rad\,yr}^{-1}$. Also other permitted values exist for $a^\ast$ and $r_0$, but they are not considered here since they correspond to smaller figures for the spin parameter which are not compatible with the majority of the constraints on it existing in the literature.

Furthermore, the LT prediction $\psi^\mathrm{LT}_\mathrm{jet} = 1.16^\circ$ for the angle  between $\boldsymbol{\hat{k}}$ and $\boldsymbol{\hat{h}}$, calculated for the aforementioned values of $a^\ast$ and $r_0$, agrees with its measured counterpart $\psi_\mathrm{jet}^\mathrm{exp} = 1.25^\circ\pm 0.18^\circ$.

Finally, the  time series for the angles  $\phi\ton{t}$ and $\eta\ton{t}$ which determine the orientation of the disk's angular momentum in space, obtained by numerically integrating the equations for their LT rates of change over some decades, are able to reproduce the measured ones both qualitatively and quantitatively. This fact implies, among other things, that
the hypothesis of a tight jet--disk coupling receives a strong support as well.

\section*{Data availability}
No new data were generated or analysed in support of this research.
\section*{Conflict of interest statement}
I declare no conflicts of interest.
\section*{Acknowledgements}
I am grateful to M. Zaja\v{c}ek and Cui Y. for useful information and explanations. \textcolor{black}{I wish to thank also an anonymous referee for her/his important critical remarks}.
\bibliography{Megabib}{}

\providecommand{\href}[2]{#2}\begingroup\raggedright\begin{thebibliography}{100}

\bibitem{2007GReGr..39.1735P}
H.~{Pfister}, ``{On the history of the so--called Lense--Thirring effect},''
  \href{http://dx.doi.org/10.1007/s10714-007-0521-4}{{\em Gen. Relativ.
  Gravit.} {\bfseries 39} (2007) 1735--1748}.

\bibitem{2008mgm..conf.2456P}
H.~{Pfister}, \href{http://dx.doi.org/10.1142/9789812834300\textunderscore
  0433}{``{The History of the So--Called Lense--Thirring Effect},''} in {\em
  The Eleventh Marcel Grossmann Meeting On Recent Developments in Theoretical
  and Experimental General Relativity, Gravitation and Relativistic Field
  Theories}, H.~{Kleinert}, R.~T. {Jantzen}, and R.~{Ruffini}, eds.,
  pp.~2456--2458.
\newblock World Scientific, 2008.

\bibitem{Pfister2014}
H.~{Pfister}, \href{http://dx.doi.org/10.1007/978-3-319-06761-2\textunderscore
  24}{``{Gravitomagnetism: From Einstein's 1912 Paper to the Satellites LAGEOS
  and Gravity Probe B},''} in {\em Relativity and Gravitation},
  J.~{Bi\v{c}\'{a}k} and T.~{Ledvinka}, eds., vol.~157 of {\em Springer
  Proceedings in Physics,}, pp.~191--197.
\newblock Springer, 2014.

\bibitem{1918PhyZ...19..156L}
J.~{Lense} and H.~{Thirring}, ``{{\"U}ber den Einflu{\ss} der Eigenrotation der
  Zentralk{\"o}rper auf die Bewegung der Planeten und Monde nach der
  Einsteinschen Gravitationstheorie},'' {\em Phys. Z} {\bfseries 19} (1918)
  156--163.

\bibitem{1984GReGr..16..711M}
B.~{Mashhoon}, F.~W. {Hehl}, and D.~S. {Theiss}, ``{On the gravitational
  effects of rotating masses: The Thirring--Lense papers.},''
  \href{http://dx.doi.org/10.1007/BF00762913}{{\em Gen. Relativ. Gravit.}
  {\bfseries 16} (1984) 711--750}.

\bibitem{2019ApJ...875L...1E}
{Event Horizon Telescope Collaboration}, ``{First M87 Event Horizon Telescope
  Results. I. The Shadow of the Supermassive Black Hole},''
  \href{http://dx.doi.org/10.3847/2041-8213/ab0ec7}{{\em Astrophys. J. Lett.}
  {\bfseries 875} (2019) L1}, \href{http://arxiv.org/abs/1906.11238}{{\ttfamily
  arXiv:1906.11238 [astro-ph.GA]}}.

\bibitem{1918PLicO..13....9C}
H.~D. {Curtis}, ``{Descriptions of 762 Nebulae and Clusters Photographed with
  the Crossley Reflector},'' {\em Publications of Lick Observatory} {\bfseries
  13} (1918) 9--42.

\bibitem{Berman15}
B.~{Berman}, ``{Weird Object: M87},'' {\em Astronomy} (2015) .
  \url{https://www.astronomy.com/science/weird-object-m87/}.

\bibitem{2013ARA&A..51..511K}
J.~{Kormendy} and L.~C. {Ho}, ``{Coevolution (Or Not) of Supermassive Black
  Holes and Host Galaxies},''
  \href{http://dx.doi.org/10.1146/annurev-astro-082708-101811}{{\em Annu. Rev.
  Astron. Astr.} {\bfseries 51} (2013) 511--653},
  \href{http://arxiv.org/abs/1304.7762}{{\ttfamily arXiv:1304.7762
  [astro-ph.CO]}}.

\bibitem{2024MNRAS.527.2341S}
D.~A. {Simon}, M.~{Cappellari}, and J.~{Hartke}, ``{Supermassive black hole
  mass in the massive elliptical galaxy M87 from integral-field stellar
  dynamics using OASIS and MUSE with adaptive optics: assessing systematic
  uncertainties},'' \href{http://dx.doi.org/10.1093/mnras/stad3309}{{\em Mon.
  Not. Roy. Astron. Soc.} {\bfseries 527} (2024) 2341--2361},
  \href{http://arxiv.org/abs/2303.18229}{{\ttfamily arXiv:2303.18229
  [astro-ph.GA]}}.

\bibitem{2017Galax...5....2H}
K.~{Hada}, ``{The Structure and Propagation of the Misaligned Jet M87},''
  \href{http://dx.doi.org/10.3390/galaxies5010002}{{\em Galaxies} {\bfseries 5}
  (2017) 2}.

\bibitem{2023Natur.621..711C}
Y.~{Cui}, K.~{Hada}, T.~{Kawashima}, {\em et~al.}, ``{Precessing jet nozzle
  connecting to a spinning black hole in M87},''
  \href{http://dx.doi.org/10.1038/s41586-023-06479-6}{{\em Nature} {\bfseries
  621} (2023) 711--715}, \href{http://arxiv.org/abs/2310.09015}{{\ttfamily
  arXiv:2310.09015 [astro-ph.HE]}}.

\bibitem{1997ApJ...476..221B}
F.~{Banyuls}, J.~A. {Font}, J.~M. {Ib{\'a}{\~n}ez}, J.~M. {Mart{\'\i}}, and
  J.~A. {Miralles}, ``{Numerical \{3 + 1\} General Relativistic Hydrodynamics:
  A Local Characteristic Approach},''
  \href{http://dx.doi.org/10.1086/303604}{{\em Astrophys. J.} {\bfseries 476}
  (1997) 221--231}.

\bibitem{1999ApJ...522..727K}
S.~{Koide}, K.~{Shibata}, and T.~{Kudoh}, ``{Relativistic Jet Formation from
  Black Hole Magnetized Accretion Disks: Method, Tests, and Applications of a
  General RelativisticMagnetohydrodynamic Numerical Code},''
  \href{http://dx.doi.org/10.1086/307667}{{\em Astrophys. J.} {\bfseries 522}
  (1999) 727--752}.

\bibitem{2003ApJ...599.1238D}
J.-P. {De Villiers}, J.~F. {Hawley}, and J.~H. {Krolik}, ``{Magnetically Driven
  Accretion Flows in the Kerr Metric. I. Models and Overall Structure},''
  \href{http://dx.doi.org/10.1086/379509}{{\em Astrophys. J.} {\bfseries 599}
  (2003) 1238--1253}, \href{http://arxiv.org/abs/astro-ph/0307260}{{\ttfamily
  arXiv:astro-ph/0307260 [astro-ph]}}.

\bibitem{2003ApJ...589..444G}
C.~F. {Gammie}, J.~C. {McKinney}, and G.~{T{\'o}th}, ``{HARM: A Numerical
  Scheme for General Relativistic Magnetohydrodynamics},''
  \href{http://dx.doi.org/10.1086/374594}{{\em Astrophys. J.} {\bfseries 589}
  (2003) 444--457}, \href{http://arxiv.org/abs/astro-ph/0301509}{{\ttfamily
  arXiv:astro-ph/0301509 [astro-ph]}}.

\bibitem{2005PhRvD..72d4014S}
M.~{Shibata} and Y.-I. {Sekiguchi}, ``{Magnetohydrodynamics in full general
  relativity: Formulation and tests},''
  \href{http://dx.doi.org/10.1103/PhysRevD.72.044014}{{\em Phys. Rev. D}
  {\bfseries 72} (2005) 044014},
  \href{http://arxiv.org/abs/astro-ph/0507383}{{\ttfamily
  arXiv:astro-ph/0507383 [astro-ph]}}.

\bibitem{2006EAS....21...43G}
E.~{Gourgoulhon}, \href{http://dx.doi.org/10.1051/eas:2006106}{``{An
  introduction to relativistic hydrodynamics},''} in {\em EAS Publications
  Series}, M.~{Rieutord} and B.~{Dubrulle}, eds., vol.~21 of {\em EAS
  Publications Series}, pp.~43--79.
\newblock 2006.
\newblock \href{http://arxiv.org/abs/gr-qc/0603009}{{\ttfamily
  arXiv:gr-qc/0603009 [gr-qc]}}.

\bibitem{2013rehy.book.....R}
L.~{Rezzolla} and O.~{Zanotti},
  \href{http://dx.doi.org/10.1093/acprof:oso/9780198528906.001.0001}{{\em
  {Relativistic Hydrodynamics}}}.
\newblock Oxford University Press, 2013.

\bibitem{2024arXiv240413824M}
Y.~{Mizuno} and L.~{Rezzolla}, ``{General-Relativistic Magnetohydrodynamic
  Equations: the bare essential},''
  \href{http://dx.doi.org/10.48550/arXiv.2404.13824}{{\em arXiv e-prints}
  (2024) arXiv:2404.13824}, \href{http://arxiv.org/abs/2404.13824}{{\ttfamily
  arXiv:2404.13824 [astro-ph.HE]}}.

\bibitem{1965JMP.....6..915N}
E.~T. {Newman} and A.~I. {Janis}, ``{Note on the Kerr Spinning-Particle
  Metric},'' \href{http://dx.doi.org/10.1063/1.1704350}{{\em J. Math. Phys.}
  {\bfseries 6} (1965) 915--917}.

\bibitem{1965JMP.....6..918N}
E.~T. {Newman}, E.~{Couch}, K.~{Chinnapared}, {\em et~al.}, ``{Metric of a
  Rotating, Charged Mass},'' \href{http://dx.doi.org/10.1063/1.1704351}{{\em J.
  Math. Phys.} {\bfseries 6} (1965) 918--919}.

\bibitem{2024arXiv241107481M}
X.-C. {Meng}, C.-H. {Wang}, and S.-W. {Wei}, ``{Imprints of black hole charge
  on the precessing jet nozzle of M87*},'' {\em arXiv e-prints} (2024)
  arXiv:2411.07481, \href{http://arxiv.org/abs/2411.07481}{{\ttfamily
  arXiv:2411.07481 [gr-qc]}}.

\bibitem{1970Natur.226...64B}
J.~M. {Bardeen}, ``{Kerr Metric Black Holes},''
  \href{http://dx.doi.org/10.1038/226064a0}{{\em Nature} {\bfseries 226}
  no.~5240, (1970) 64--65}.

\bibitem{1963PhRvL..11..237K}
R.~P. {Kerr}, ``{Gravitational Field of a Spinning Mass as an Example of
  Algebraically Special Metrics},''
  \href{http://dx.doi.org/10.1103/PhysRevLett.11.237}{{\em Phys. Rev. Lett.}
  {\bfseries 11} (1963) 237--238}.

\bibitem{2015CQGra..32l4006T}
S.~A. {Teukolsky}, ``{The Kerr metric},''
  \href{http://dx.doi.org/10.1088/0264-9381/32/12/124006}{{\em Class. Quantum
  Gravit.} {\bfseries 32} (2015) 124006},
  \href{http://arxiv.org/abs/1410.2130}{{\ttfamily arXiv:1410.2130 [gr-qc]}}.

\bibitem{1986bhwd.book.....S}
S.~L. {Shapiro} and S.~A. {Teukolsky}, {\em {Black Holes, White Dwarfs and
  Neutron Stars: The Physics of Compact Objects}}.
\newblock Wiley, 1986.

\bibitem{1973CMaPh..34..135Y}
P.~{Yodzis}, H.-J. {Seifert}, and H.~{M{\"u}ller Zum Hagen}, ``{On the
  occurrence of naked singularities in general relativity},''
  \href{http://dx.doi.org/10.1007/BF01646443}{{\em Communications in
  Mathematical Physics} {\bfseries 34} (1973) 135--148}.

\bibitem{1991PhRvL..66..994S}
S.~L. {Shapiro} and S.~A. {Teukolsky}, ``{Formation of naked singularities: The
  violation of cosmic censorship},''
  \href{http://dx.doi.org/10.1103/PhysRevLett.66.994}{{\em Phys. Rev. Lett.}
  {\bfseries 66} (1991) 994--997}.

\bibitem{1999JApA...20..233P}
R.~{Penrose}, ``{The Question of Cosmic Censorship},''
  \href{http://dx.doi.org/10.1007/BF02702355}{{\em J. Astrophys. Astron.}
  {\bfseries 20} (1999) 233--248}.

\bibitem{2002GReGr..34.1141P}
R.~{Penrose}, ``{``Golden Oldie'': Gravitational Collapse: The Role of General
  Relativity},'' \href{http://dx.doi.org/10.1023/A:1016578408204}{{\em Gen.
  Relativ. Gravit.} {\bfseries 7} (2002) 1141--1165}.

\bibitem{1972ApJ...178..347B}
J.~M. {Bardeen}, W.~H. {Press}, and S.~A. {Teukolsky}, ``{Rotating Black Holes:
  Locally Nonrotating Frames, Energy Extraction, and Scalar Synchrotron
  Radiation},'' \href{http://dx.doi.org/10.1086/151796}{{\em Astrophys. J.}
  {\bfseries 178} (1972) 347--370}.

\bibitem{1972PhRvD...5..814W}
D.~C. {Wilkins}, ``{Bound Geodesics in the Kerr Metric},''
  \href{http://dx.doi.org/10.1103/PhysRevD.5.814}{{\em Phys. Rev. D} {\bfseries
  5} (1972) 814--822}.

\bibitem{2024PhRvD.109b4029A}
A.~M. {Al Zahrani}, ``{Tilted circular orbits around a Kerr black hole},''
  \href{http://dx.doi.org/10.1103/PhysRevD.109.024029}{{\em Phys. Rev. D}
  {\bfseries 109} (2024) 024029},
  \href{http://arxiv.org/abs/2312.12988}{{\ttfamily arXiv:2312.12988 [gr-qc]}}.

\bibitem{2024arXiv240602454G}
R.~{Ghosh} and K.~{Chakravarti}, ``{Parameterized Non-circular Deviation from
  the Kerr Paradigm and Its Observational Signatures: Extreme Mass Ratio
  Inspirals and Lense-Thirring Effect},''
  \href{http://dx.doi.org/10.48550/arXiv.2406.02454}{{\em arXiv e-prints}
  (2024) arXiv:2406.02454}, \href{http://arxiv.org/abs/2406.02454}{{\ttfamily
  arXiv:2406.02454 [gr-qc]}}.

\bibitem{2009ApJ...692.1075G}
S.~{Gillessen}, F.~{Eisenhauer}, S.~{Trippe}, {\em et~al.}, ``{Monitoring
  Stellar Orbits Around the Massive Black Hole in the Galactic Center},''
  \href{http://dx.doi.org/10.1088/0004-637X/692/2/1075}{{\em Astrophys. J.}
  {\bfseries 692} (2009) 1075--1109},
  \href{http://arxiv.org/abs/0810.4674}{{\ttfamily arXiv:0810.4674
  [astro-ph]}}.

\bibitem{2010RvMP...82.3121G}
R.~{Genzel}, F.~{Eisenhauer}, and S.~{Gillessen}, ``{The Galactic Center
  massive black hole and nuclear star cluster},''
  \href{http://dx.doi.org/10.1103/RevModPhys.82.3121}{{\em Rev. Mod. Phys.}
  {\bfseries 82} (2010) 3121--3195},
  \href{http://arxiv.org/abs/1006.0064}{{\ttfamily arXiv:1006.0064
  [astro-ph.GA]}}.

\bibitem{2017ApJ...837...30G}
S.~{Gillessen}, P.~M. {Plewa}, F.~{Eisenhauer}, {\em et~al.}, ``{An Update on
  Monitoring Stellar Orbits in the Galactic Center},''
  \href{http://dx.doi.org/10.3847/1538-4357/aa5c41}{{\em Astrophys. J.}
  {\bfseries 837} (2017) 30}, \href{http://arxiv.org/abs/1611.09144}{{\ttfamily
  arXiv:1611.09144 [astro-ph.GA]}}.

\bibitem{Thorne86}
K.~S. {Thorne}, D.~A. {MacDonald}, and R.~H. {Price}, eds., {\em {Black Holes:
  The Membrane Paradigm}}.
\newblock Yale University Press, 1986.

\bibitem{1986hmac.book..103T}
K.~S. {Thorne}, ``{Black Holes: The Membrane Viewpoint},'' in {\em Highlights
  of Modern Astrophysics: Concepts and Controversies}, S.~L. {Shapiro}, S.~A.
  {Teukolsky}, and E.~E. {Salpeter}, eds., pp.~103--161.
\newblock Wiley, 1986.

\bibitem{1988nznf.conf..573T}
K.~S. {Thorne}, ``{Gravitomagnetism, jets in quasars, and the Stanford
  Gyroscope Experiment.},'' in {\em Near Zero: New Frontiers of Physics}, J.~D.
  {Fairbank}, J.~{Deaver}, B.~S., C.~W.~F. {Everitt}, and P.~F. {Michelson},
  eds., pp.~573--586.
\newblock Freeman, 1988.

\bibitem{2001rfg..conf..121M}
B.~{Mashhoon}, ``{Gravitoelectromagnetism},'' in {\em Reference Frames and
  Gravitomagnetism}, J.~F. {Pascual-S{\'a}nchez}, L.~{Flor{\'\i}a}, A.~{San
  Miguel}, and F.~{Vicente}, eds.
\newblock World Scientific, 2001.

\bibitem{2001rsgc.book.....R}
W.~{Rindler}, {\em {Relativity: special, general, and cosmological}}.
\newblock Oxford University Press, 2001.

\bibitem{Mash07}
B.~{Mashhoon}, ``{Gravitoelectromagnetism: A Brief Review},'' in {\em The
  Measurement of Gravitomagnetism: A Challenging Enterprise}, L.~{Iorio}, ed.,
  pp.~29--39.
\newblock Nova Science, 2007.

\bibitem{2024gpno.book.....I}
L.~{Iorio}, \href{http://dx.doi.org/10.1017/9781009562911}{{\em {General
  Post-Newtonian Orbital Effects From Earth's Satellites to the Galactic
  Center}}}.
\newblock Cambridge University Press, 2024.

\bibitem{2019JGeod..93.2181P}
M.~{Pearlman}, D.~{Arnold}, M.~{Davis}, {\em et~al.}, ``{Laser geodetic
  satellites: a high-accuracy scientific tool},''
  \href{http://dx.doi.org/10.1007/s00190-019-01228-y}{{\em J. Geod.} {\bfseries
  93} (2019) 2181--2194}.

\bibitem{SLR11}
D.~{Coulot}, F.~{Deleflie}, P.~{Bonnefond}, {\em et~al.},
  \href{http://dx.doi.org/10.1007/978-90-481-8702-7\textunderscore
  98}{``Satellite laser ranging,''} in {\em Encyclopedia of Solid Earth
  Geophysics}, H.~K. {Gupta}, ed., Encyclopedia of Earth Sciences Series,
  pp.~1049--1055.
\newblock Springer, 2011.

\bibitem{1996NCimA.109..575C}
I.~{Ciufolini}, D.~M. {Lucchesi}, F.~{Vespe}, and A.~{Mandiello},
  ``{Measurement of dragging of inertial frames and gravitomagnetic field using
  laser--ranged satellites.},''
  \href{http://dx.doi.org/10.1007/BF02731140}{{\em Nuovo Cim. A} {\bfseries
  109A} (1996) 575--590}.

\bibitem{2011Ap&SS.331..351I}
L.~{Iorio}, H.~I.~M. {Lichtenegger}, M.~L. {Ruggiero}, and C.~{Corda},
  ``{Phenomenology of the Lense--Thirring effect in the solar system},''
  \href{http://dx.doi.org/10.1007/s10509-010-0489-5}{{\em Astrophys. Space
  Sci.} {\bfseries 331} (2011) 351--395},
  \href{http://arxiv.org/abs/1009.3225}{{\ttfamily arXiv:1009.3225 [gr-qc]}}.

\bibitem{2013NuPhS.243..180C}
I.~{Ciufolini}, A.~{Paolozzi}, R.~{Koenig}, {\em et~al.}, ``{Fundamental
  Physics and General Relativity with the LARES and LAGEOS satellites},''
  \href{http://dx.doi.org/10.1016/j.nuclphysbps.2013.09.005}{{\em Nucl. Phys. B
  Proc. Suppl.} {\bfseries 243} (2013) 180--193},
  \href{http://arxiv.org/abs/1309.1699}{{\ttfamily arXiv:1309.1699 [gr-qc]}}.

\bibitem{2013CEJPh..11..531R}
G.~{Renzetti}, ``{History of the attempts to measure orbital frame--dragging
  with artificial satellites},''
  \href{http://dx.doi.org/10.2478/s11534-013-0189-1}{{\em Centr. Eur. J. Phys.}
  {\bfseries 11} (2013) 531--544}.

\bibitem{2011AGUFM.P41B1620F}
S.~{Finocchiaro}, L.~{Iess}, W.~M. {Folkner}, and S.~{Asmar}, ``{The
  Determination of Jupiter's Angular Momentum from the Lense-Thirring
  Precession of the Juno Spacecraft},'' in {\em AGU Fall Meeting Abstracts},
  vol.~2011, pp.~P41B--1620.
\newblock 2011.

\bibitem{2024ApJ...971..145D}
D.~{Durante}, P.~{Cappuccio}, I.~{di Stefano}, {\em et~al.}, ``{Testing General
  Relativity with Juno at Jupiter},''
  \href{http://dx.doi.org/10.3847/1538-4357/ad5ff5}{{\em Astrophys. J.}
  {\bfseries 971} (2024) 145}.

\bibitem{2015IAUGA..2227771P}
R.~S. {Park}, W.~M. {Folkner}, and A.~S. {Konopliv}, ``{Estimation of Solar
  Angular Momentum from Lense--Thirring Precession of Mercury},'' {\em IAU
  General Assembly} {\bfseries 22} (2015) 2227771.

\bibitem{Pav2024}
D.~{Pavlov} and I.~{Dolgakov}, ``{General relativity tests by the dynamics of
  the Solar system},'' in {\em Proceedings of the Journ\'{e}es 2023
  \virg{Syst\`{e}mes de rf\'{e}r\'{e}nce spatio-temporels}}, C.~{Bizouard},
  A.~{Fienga}, and F.~{Paganelli}, eds., pp.~156--160.
\newblock Observatoire de la C\^{o}te d'Azur, 2024.

\bibitem{RussiLT}
D.~{Pavlov} and I.~{Dolgakov}, ``{Studying the Properties of Spacetime with an
  Improved Dynamical Model of the Inner Solar System},''
  \href{http://dx.doi.org/10.3390/universe10110413}{{\em Universe} {\bfseries
  10} (2024) 413}.

\bibitem{Varenna74}
C.~W.~F. {Everitt}, ``{The Gyroscope experiment - I: General description and
  analysis of gyroscope performance},'' in {\em Proceedings of the
  International School of Physics \virg{Enrico Fermi}. Course LVI. Experimental
  Gravitation}, B.~{Bertotti}, ed., pp.~331--360.
\newblock Academic Press, 1974.

\bibitem{2001LNP...562...52E}
C.~W.~F. {Everitt}, S.~{Buchman}, D.~B. {Debra}, {\em et~al.},
  \href{http://dx.doi.org/10.1007/3-540-40988-2\textunderscore 4}{``{Gravity
  Probe B: Countdown to Launch},''} in {\em Gyros, Clocks, Interferometers ...:
  Testing Relativistic Gravity in Space}, C.~{L{\"a}mmerzahl}, C.~W.~F.
  {Everitt}, and F.~W. {Hehl}, eds., vol.~562 of {\em Lecture Notes in
  Physics}, pp.~52--82.
\newblock Springer, 2001.

\bibitem{Pugh59}
G.~E. {Pugh}, ``{Proposal for a Satellite Test of the Coriolis Prediction of
  General Relativity},'' research memorandum, Weapons Systems Evaluation Group,
  The Pentagon, Washington D.C., 1959.

\bibitem{Schiff60}
L.~{Schiff}, ``{Possible new experimental test of general relativity theory},''
  \href{http://dx.doi.org/10.1103/PhysRevLett.4.215}{{\em Phys. Rev. Lett.}
  {\bfseries 4} (1960) 215--217}.

\bibitem{2011PhRvL.106v1101E}
C.~W.~F. {Everitt}, D.~B. {Debra}, B.~W. {Parkinson}, {\em et~al.}, ``{Gravity
  Probe B: Final Results of a Space Experiment to Test General Relativity},''
  \href{http://dx.doi.org/10.1103/PhysRevLett.106.221101}{{\em Phys. Rev.
  Lett.} {\bfseries 106} (2011) 221101},
  \href{http://arxiv.org/abs/1105.3456}{{\ttfamily arXiv:1105.3456 [gr-qc]}}.

\bibitem{2015CQGra..32v4001E}
C.~W.~F. {Everitt}, B.~{Muhlfelder}, D.~B. {Debra}, {\em et~al.}, ``{The
  Gravity Probe B test of general relativity},''
  \href{http://dx.doi.org/10.1088/0264-9381/32/22/224001}{{\em Class. Quantum
  Gravit.} {\bfseries 32} (2015) 224001}.

\bibitem{Sof89}
M.~H. {Soffel}, \href{http://dx.doi.org/10.1007/978-3-642-73406-9}{{\em
  Relativity in Astrometry, Celestial Mechanics and Geodesy}}.
\newblock Springer, 1989.

\bibitem{1991ercm.book.....B}
V.~A. {Brumberg}, {\em {Essential Relativistic Celestial Mechanics}}.
\newblock Adam Hilger, 1991.

\bibitem{2000ssd..book.....M}
C.~D. {Murray} and S.~F. {Dermott},
  \href{http://dx.doi.org/10.1017/CBO9781139174817}{{\em {Solar System
  Dynamics}}}.
\newblock Cambridge University Press, 1999.

\bibitem{2003ASSL..293.....B}
B.~{Bertotti}, P.~{Farinella}, and D.~{Vokrouhlick\'{y}},
  \href{http://dx.doi.org/10.1007/978-94-010-0233-2}{{\em {Physics of the Solar
  System}}}.
\newblock Kluwer, 2003.

\bibitem{2005ormo.book.....R}
A.~E. {Roy}, {\em {Orbital Motion. Fourth Edition}}.
\newblock IOP Publishing, 2005.

\bibitem{2011rcms.book.....K}
S.~M. {Kopeikin}, M.~{Efroimsky}, and G.~{Kaplan},
  \href{http://dx.doi.org/10.1002/9783527634569}{{\em {Relativistic Celestial
  Mechanics of the Solar System}}}.
\newblock Wiley, 2011.

\bibitem{2014grav.book.....P}
E.~{Poisson} and C.~M. {Will},
  \href{http://dx.doi.org/10.1017/CBO9781139507486}{{\em {Gravity. Newtonian,
  Post--Newtonian, Relativistic}}}.
\newblock Cambridge University Press, 2014.

\bibitem{SoffelHan19}
M.~H. {Soffel} and W.-B. {Han},
  \href{http://dx.doi.org/10.1007/978-3-030-19673-8}{{\em {Applied General
  Relativity}}}.
\newblock {Astronomy and Astrophysics Library}. Springer, 2019.

\bibitem{Gold80}
H.~{Goldstein}, {\em {Classical Mechanics. Second Edition}}.
\newblock Addison Wesley, 1980.

\bibitem{Taff85}
L.~G. {Taff}, {\em {Celestial Mechanics: A Computational Guide for the
  Practitioner}}.
\newblock Wiley, 1985.

\bibitem{1974PhRvD..10.1340B}
B.~M. {Barker} and R.~F. {Oconnell}, ``{Effect of the rotation of the central
  body on the orbit of a satellite},''
  \href{http://dx.doi.org/10.1103/PhysRevD.10.1340}{{\em Phys. Rev. D}
  {\bfseries 10} (1974) 1340--1342}.

\bibitem{2013Sci...339...49M}
J.~{McKinney}, A.~{Tchekhovskoy}, and R.~D. {Blandford}, ``{Alignment of
  Magnetized Accretion Disks and Relativistic Jets with Spinning Black
  Holes},'' \href{http://dx.doi.org/10.1126/science.1230811}{{\em Science}
  {\bfseries 339} (2013) 49}, \href{http://arxiv.org/abs/1211.3651}{{\ttfamily
  arXiv:1211.3651 [astro-ph.CO]}}.

\bibitem{2018MNRAS.474L..81L}
M.~{Liska}, C.~{Hesp}, A.~{Tchekhovskoy}, {\em et~al.}, ``{Formation of
  precessing jets by tilted black hole discs in 3D general relativistic MHD
  simulations},'' \href{http://dx.doi.org/10.1093/mnrasl/slx174}{{\em Mon. Not.
  Roy. Astron. Soc.} {\bfseries 474} (2018) L81--L85},
  \href{http://arxiv.org/abs/1707.06619}{{\ttfamily arXiv:1707.06619
  [astro-ph.HE]}}.

\bibitem{2024arXiv241010965C}
Y.~{Cui} and W.~{Lin}, ``{Imprints of M87 Jet Precession on the Black
  Hole-Accretion Disk System},''
  \href{http://dx.doi.org/10.48550/arXiv.2410.10965}{{\em arXiv e-prints}
  (2024) arXiv:2410.10965}, \href{http://arxiv.org/abs/2410.10965}{{\ttfamily
  arXiv:2410.10965 [astro-ph.HE]}}.

\bibitem{2023ApJ...951..106B}
S.~{Britzen}, M.~{Zaja{\v{c}}ek}, {Gopal-Krishna}, {\em et~al.},
  ``{Precession-induced Variability in AGN Jets and OJ 287},''
  \href{http://dx.doi.org/10.3847/1538-4357/accbbc}{{\em Astrophys. J.}
  {\bfseries 951} (2023) 106},
  \href{http://arxiv.org/abs/2307.05838}{{\ttfamily arXiv:2307.05838
  [astro-ph.HE]}}.

\bibitem{2024Natur.630..325P}
D.~R. {Pasham}, M.~{Zaja{\v{c}}ek}, C.~J. {Nixon}, {\em et~al.},
  ``{Lense-Thirring precession after a supermassive black hole disrupts a
  star},'' \href{http://dx.doi.org/10.1038/s41586-024-07433-w}{{\em Nature}
  {\bfseries 630} (2024) 325--328},
  \href{http://arxiv.org/abs/2402.09689}{{\ttfamily arXiv:2402.09689
  [astro-ph.HE]}}.

\bibitem{1982MNRAS.199..883B}
R.~D. {Blandford} and D.~G. {Payne}, ``{Hydromagnetic flows from accretion
  disks and the production of radio jets.},''
  \href{http://dx.doi.org/10.1093/mnras/199.4.883}{{\em Mon. Not. Roy. Astron.
  Soc.} {\bfseries 199} (1982) 883--903}.

\bibitem{1977MNRAS.179..433B}
R.~D. {Blandford} and R.~L. {Znajek}, ``{Electromagnetic extraction of energy
  from Kerr black holes.},''
  \href{http://dx.doi.org/10.1093/mnras/179.3.433}{{\em Mon. Not. Roy. Astron.
  Soc.} {\bfseries 179} (1977) 433--456}.

\bibitem{2024ApJ...973..141G}
M.~{Guo}, J.~M. {Stone}, E.~{Quataert}, and C.-G. {Kim}, ``{Magnetized
  Accretion onto and Feedback from Supermassive Black Holes in Elliptical
  Galaxies},'' \href{http://dx.doi.org/10.3847/1538-4357/ad5fe7}{{\em
  Astrophys. J.} {\bfseries 973} (2024) 141},
  \href{http://arxiv.org/abs/2405.11711}{{\ttfamily arXiv:2405.11711
  [astro-ph.HE]}}.

\bibitem{2023Galax..11....4A}
R.~{Anantua}, J.~{D{\'u}ran}, N.~{Ngata}, L.~{Oramas}, J.~{R{\"o}der},
  R.~{Emami}, A.~{Ricarte}, B.~{Curd}, A.~E. {Broderick}, J.~{Wayland}, G.~N.
  {Wong}, S.~{Ressler}, N.~{Nigam}, and E.~{Durodola}, ``{Emission Modeling in
  the EHT{\textendash}ngEHT Age},''
  \href{http://dx.doi.org/10.3390/galaxies11010004}{{\em Galaxies} {\bfseries
  11} (2023) 4}, \href{http://arxiv.org/abs/2211.06541}{{\ttfamily
  arXiv:2211.06541 [astro-ph.HE]}}.

\bibitem{2024ApJ...974..209G}
S.~{Gupta} and J.~{Dexter}, ``{Shock-induced Partial Alignment in Geometrically
  Thick Tilted Accretion Disks Around Black Holes},''
  \href{http://dx.doi.org/10.3847/1538-4357/ad737d}{{\em Astrophys. J.}
  {\bfseries 974} (2024) 209},
  \href{http://arxiv.org/abs/2409.09165}{{\ttfamily arXiv:2409.09165
  [astro-ph.HE]}}.

\bibitem{1998ApJ...492L..59S}
L.~{Stella} and M.~{Vietri}, ``{Lense--Thirring Precession and Quasi--periodic
  Oscillations in Low-Mass X--Ray Binaries},''
  \href{http://dx.doi.org/10.1086/311075}{{\em Astrophys. J. Lett.} {\bfseries
  492} (1998) L59--L62},
  \href{http://arxiv.org/abs/astro-ph/9709085}{{\ttfamily
  arXiv:astro-ph/9709085 [astro-ph]}}.

\bibitem{1999ApJ...524L..63S}
L.~{Stella}, M.~{Vietri}, and S.~M. {Morsink}, ``{Correlations in the
  Quasi-periodic Oscillation Frequencies of Low-Mass X-Ray Binaries and the
  Relativistic Precession Model},''
  \href{http://dx.doi.org/10.1086/312291}{{\em Astrophys. J. Lett.} {\bfseries
  524} (1999) L63--L66},
  \href{http://arxiv.org/abs/astro-ph/9907346}{{\ttfamily
  arXiv:astro-ph/9907346 [astro-ph]}}.

\bibitem{2009MNRAS.397L.101I}
A.~{Ingram}, C.~{Done}, and P.~C. {Fragile}, ``{Low--frequency quasi---periodic
  oscillations spectra and Lense--Thirring precession},''
  \href{http://dx.doi.org/10.1111/j.1745-3933.2009.00693.x}{{\em Mon. Not. Roy.
  Astron. Soc.} {\bfseries 397} (2009) L101--L105},
  \href{http://arxiv.org/abs/0901.1238}{{\ttfamily arXiv:0901.1238
  [astro-ph.SR]}}.

\bibitem{2016AN....337..398M}
S.~E. {Motta}, ``{Quasi periodic oscillations in black hole binaries},''
  \href{http://dx.doi.org/10.1002/asna.201612320}{{\em Astron. Nachr.}
  {\bfseries 337} (2016) 398},
  \href{http://arxiv.org/abs/1603.07885}{{\ttfamily arXiv:1603.07885
  [astro-ph.HE]}}.

\bibitem{1998AJ....116.2237K}
M.~{Kissler-Patig} and K.~{Gebhardt}, ``{The Spin of M87 as Measured from the
  Rotation of its Globular Clusters},''
  \href{http://dx.doi.org/10.1086/300609}{{\em Astron J.} {\bfseries 116}
  (1998) 2237--2245}, \href{http://arxiv.org/abs/astro-ph/9807231}{{\ttfamily
  arXiv:astro-ph/9807231 [astro-ph]}}.

\bibitem{2008ApJ...676L.109W}
J.-M. {Wang}, Y.-R. {Li}, J.-C. {Wang}, and S.~{Zhang}, ``{Spins of the
  Supermassive Black Hole in M87: New Constraints from TeV Observations},''
  \href{http://dx.doi.org/10.1086/587740}{{\em Astrophys. J. Lett.} {\bfseries
  676} (2008) L109}, \href{http://arxiv.org/abs/0802.4322}{{\ttfamily
  arXiv:0802.4322 [astro-ph]}}.

\bibitem{2009ApJ...699..513L}
Y.-R. {Li}, Y.-F. {Yuan}, J.-M. {Wang}, {\em et~al.}, ``{Spins of Supermassive
  Black Holes in M87. II. Fully General Relativistic Calculations},''
  \href{http://dx.doi.org/10.1088/0004-637X/699/1/513}{{\em Astrophys. J.}
  {\bfseries 699} (2009) 513--524},
  \href{http://arxiv.org/abs/0904.2335}{{\ttfamily arXiv:0904.2335
  [astro-ph.HE]}}.

\bibitem{2012Sci...338..355D}
S.~S. {Doeleman}, V.~L. {Fish}, D.~E. {Schenck}, {\em et~al.}, ``{Jet-Launching
  Structure Resolved Near the Supermassive Black Hole in M87},''
  \href{http://dx.doi.org/10.1126/science.1224768}{{\em Science} {\bfseries
  338} (2012) 355}, \href{http://arxiv.org/abs/1210.6132}{{\ttfamily
  arXiv:1210.6132 [astro-ph.HE]}}.

\bibitem{2017MNRAS.470..612F}
J.~{Feng} and Q.~{Wu}, ``{Constraint on the black hole spin of M87 from the
  accretion-jet model},'' \href{http://dx.doi.org/10.1093/mnras/stx1283}{{\em
  Mon. Not. Roy. Astron. Soc.} {\bfseries 470} (2017) 612--616},
  \href{http://arxiv.org/abs/1705.07804}{{\ttfamily arXiv:1705.07804
  [astro-ph.HE]}}.

\bibitem{2018MNRAS.479L..65S}
D.~N. {Sob'yanin}, ``{Black hole spin from wobbling and rotation of the M87 jet
  and a sign of a magnetically arrested disc},''
  \href{http://dx.doi.org/10.1093/mnrasl/sly097}{{\em Mon. Not. Roy. Astron.
  Soc.} {\bfseries 479} (2018) L65--L69},
  \href{http://arxiv.org/abs/1807.06296}{{\ttfamily arXiv:1807.06296
  [astro-ph.HE]}}.

\bibitem{2019ApJ...880L..26N}
R.~{Nemmen}, ``{The Spin of M87*},''
  \href{http://dx.doi.org/10.3847/2041-8213/ab2fd3}{{\em Astrophys. J. Lett.}
  {\bfseries 880} (2019) L26},
  \href{http://arxiv.org/abs/1905.02143}{{\ttfamily arXiv:1905.02143
  [astro-ph.HE]}}.

\bibitem{2019MNRAS.489.1197N}
E.~E. {Nokhrina}, L.~I. {Gurvits}, V.~S. {Beskin}, {\em et~al.}, ``{M87 black
  hole mass and spin estimate through the position of the jet boundary shape
  break},'' \href{http://dx.doi.org/10.1093/mnras/stz2116}{{\em Mon. Not. Roy.
  Astron. Soc.} {\bfseries 489} (2019) 1197--1205},
  \href{http://arxiv.org/abs/1904.05665}{{\ttfamily arXiv:1904.05665
  [astro-ph.HE]}}.

\bibitem{2020AnP...53200480G}
D.~{Garofalo}, ``{Spin of the M87 Black Hole},''
  \href{http://dx.doi.org/10.1002/andp.201900480}{{\em Ann. Phys.--Berlin}
  {\bfseries 532} (2020) 1900480},
  \href{http://arxiv.org/abs/2003.02163}{{\ttfamily arXiv:2003.02163
  [astro-ph.HE]}}.

\bibitem{2020MNRAS.492L..22T}
F.~{Tamburini}, B.~{Thid{\'e}}, and M.~{Della Valle}, ``{Measurement of the
  spin of the M87 black hole from its observed twisted light},''
  \href{http://dx.doi.org/10.1093/mnrasl/slz176}{{\em Mon. Not. Roy. Astron.
  Soc.} {\bfseries 492} (2020) L22--L27},
  \href{http://arxiv.org/abs/1904.07923}{{\ttfamily arXiv:1904.07923
  [astro-ph.HE]}}.

\bibitem{2023Astro...2..141D}
V.~I. {Dokuchaev}, ``{Spins of Supermassive Black Holes M87* and SgrA* Revealed
  from the Size of Dark Spots in Event Horizon Telescope Images},''
  \href{http://dx.doi.org/10.3390/astronomy2030010}{{\em Astronomy} {\bfseries
  2} (2023) 141--152}, \href{http://arxiv.org/abs/2307.14714}{{\ttfamily
  arXiv:2307.14714 [astro-ph.HE]}}.

\bibitem{2004ApJ...602..312G}
C.~F. {Gammie}, S.~L. {Shapiro}, and J.~C. {McKinney}, ``{Black Hole Spin
  Evolution},'' \href{http://dx.doi.org/10.1086/380996}{{\em Astrophys. J.}
  {\bfseries 602} (2004) 312--319},
  \href{http://arxiv.org/abs/astro-ph/0310886}{{\ttfamily
  arXiv:astro-ph/0310886 [astro-ph]}}.

\bibitem{2024PhRvD.110f4006W}
S.-W. {Wei}, Y.-C. {Zou}, Y.-P. {Zhang}, and Y.-X. {Liu}, ``{Constraining black
  hole parameters with the precessing jet nozzle of M87*},''
  \href{http://dx.doi.org/10.1103/PhysRevD.110.064006}{{\em Phys. Rev. D}
  {\bfseries 110} (2024) 064006},
  \href{http://arxiv.org/abs/2401.17689}{{\ttfamily arXiv:2401.17689 [gr-qc]}}.

\end{thebibliography}\endgroup
\end{document}